\documentclass[twocolumn,showpacs,showkeys,amssymb,aps,nofootinbib,floatfix,superscriptaddress]{revtex4-1}

\usepackage{epsfig}
\usepackage{hyperref}
\usepackage{comment}

\usepackage{graphicx,color}

\newcommand{\br}[1]{\langle #1\rangle}

\newcommand{\bn}[1]{\left( #1\right)}

\newcommand{\var}[1]{{\rm v}(#1)}
\newcommand{\varov}[1]{\overline{{\rm v}}(#1)}
\newcommand{\cov}[2]{{\rm c}(#1,#2)}
\newcommand{\co}[2]{{\rm C}(#1,#2)}
\newcommand{\covov}[2]{\overline{{\rm c}}(#1,#2)}

\begin{document} 
\hbadness=10000

\title{Partial correlation analysis method in ultra-relativistic heavy-ion collisions}

\author{Adam Olszewski}
\email{Adam.Olszewski.fiz@gmail.com}
\affiliation{Institute of Physics, Jan Kochanowski University, 25-406 Kielce, Poland}

\author{Wojciech Broniowski}
\email{Wojciech.Broniowski@ifj.edu.pl}
\affiliation{The H. Niewodnicza\'nski Institute of Nuclear Physics, Polish Academy of Sciences, 31-342 Cracow, Poland}
\affiliation{Institute of Physics, Jan Kochanowski University, 25-406 Kielce, Poland}

\begin{abstract}
We argue that statistical data analysis of two-particle longitudinal correlations in ultra-relativistic 
heavy-ion collisions may be efficiently carried out with the technique of partial 
covariance. In this method, the spurious event-by-event fluctuations due to imprecise centrality determination are eliminated via projecting out the component of the 
covariance influenced by the centrality fluctuations. We bring up the relationship of the partial covariance to the conditional covariance. 
Importantly, in the superposition approach, where hadrons are produced independently from a collection of sources, the framework allows us to impose
centrality constraints on the number of sources rather than hadrons, that way unfolding of the trivial fluctuations from statistical hadronization 
and focusing better on the initial-state physics.
We show, using simulated data from hydrodynamics followed with statistical hadronization, that the technique is practical and very simple to use, 
giving insight into the correlations generated in the initial stage. 
We also discuss the issues related to separation of the short- and long-range components of the correlation functions, and show that 
in our example the short-range component from the resonance decays is largely reduced by considering pions of the same sign.
We demonstrate the method explicitly on the cases where centrality is determined with a single central control bin, or with two peripheral control bins. 
\end{abstract}

\pacs{25.75.-q, 25.75Gz, 25.75.Ld}

\keywords{relativistic heavy-ion collisions, partial correlation analysis, superposition model}

\maketitle

\section{Introduction \label{sec:intro}}

From the outset of the studies of correlations in ultra-relativistic heavy-ion collisions, it has been known that 
event-by-event fluctuations due to the choice of the centrality class of the sample lead to spurious effects that should be separated
from the physical correlations.
A simple example involves the forward-backward (FB) fluctuations of the multiplicity of produced hadrons, with the centrality defined from multiplicity 
in a suitable reference bin, or via some other quantity obtained from detector response, correlated to multiplicity. As the measurements 
in the  physical $F$ and $B$  bins are correlated to the reference bin, 
the FB correlations determined via a naive definition depend strongly on the fluctuations in the reference bin, thus the character of the ``true''  FB correlations is obscured.  

Several remedies have been proposed to cure the problem of {\em centrality 
fluctuations}. Firstly, if the size of the data sample allows, 
one may use sufficiently narrow centrality bins, such that centrality fluctuations 
are negligible. Then, to improve statistics, one may average the obtained covariance matrices over several 
narrow centrality bins within a broader class. This method, essentially based on the concept of {\em conditional correlation} (see, e.g.,~\cite{Cramer:1946}), was
successfully used in the analysis by the STAR Collaboration~\cite{Abelev:2009ag}, with further proposals presented in~\cite{De:2013bta}.

In this paper we explore the so-called {\em partial correlation} analysis (see, e.g.,~\cite{Cramer:1946,krzanowski:2000}) in application to 
ultra-relativistic heavy-ion collisions. The method is closely related to the conditional correlations (see App.~\ref{sec:cond}), however, it
offers an appealing simplicity as well as immediate insight into conditional independence of the studied variables. 
It has been widely used in other 
domains of statistical applications, ranging from  physics (see, e.g., an interesting example from the X-ray 
spectroscopy, where partial correlations are used to remove the spurious effects of beam intensity fluctuation~\cite{Frasinski}) to medicine and psychology. 
The basic goal of the approach is to assess the correlation (or independence) of some ``physical'' variables, where the sample is determined by 
certain fluctuating control (or {\em external} or {\em nuisance}) variables, whose effects needs to be removed to accomplish the understanding of the relations between the physical 
variables. 

We bring up the relationship of the partial covariance to the conditional covariance~\cite{Lawrance:1976,Baba:2004}, which holds under
conditions which are well satisfied in ultra-relativistic heavy-ion collisions. Then, the partial covariance with a control bin may be understood as conditional covariance
with the fixed hadron multiplicity in the control bin. An important point, however, is that one should impose centrality constraints 
at the level of initial state rather than finally produced hadrons; that way, we limit the external fluctuations concerning the initial production, which 
are of our principal interest. One can accomplish this goal in the framework of the superposition model~\cite{Olszewski:2013qwa} of particle production, where hadrons are 
emitted from independent sources. We extend the partial correlation analysis to this physically interesting case. The result is
a simple modification of the partial-correlation formulas, where variances 
have the autocorrelation terms  removed. That way one can impose the constraints at the level of the initial sources, 
which is a non-trivial outcome of the partial covariance method
in the superposition approach.

In our study, we use simulated events for ultra-relativistic heavy-ion collisions generated with event-by-event 3+1D viscous hydrodynamics~\cite{Bozek:2011ua}
and {\tt THERMINATOR}~\cite{Kisiel:2005hn,Chojnacki:2011hb} to argue that 
the partial covariance technique is a practical tool to analyze two-particle correlations. 
We apply it to examples where centrality is determined 
with the multiplicity in a single mid-rapidity control bin, as well as in two peripheral-rapidity control bins. 

We verify in our examples that with the removed effects of resonance decays, the partial FB multiplicity correlations obtained from the simulated data 
reflect very closely the initial correlations in spatial rapidity, here implemented in the very simple Bzdak-Teaney (BT)~\cite{Bzdak:2012tp} 
form for the wounded quark model~\cite{Bozek:2016kpf}. 
This feature shows that the partial correlation technique may be used in practice to 
access information on the initial-state correlations. However, as in other approaches, this possibility relies on the separation 
of the short-range correlations (such as from the resonance decays, jets, or femtoscopic correlations) from the long-range correlations, generated in the initial 
state.  

We recall that another approach separating centrality fluctuations, somewhat different in spirit and derived from the superposition approach, defines the so-called 
{\em strongly-intensive measures}~\cite{Gazdzicki:1992ri,Gorenstein:2011vq}, which by construction do not depend on fluctuations of the number of sources (or the volume) which 
emit the observed hadrons. Yet another technique which can be used to accomplish the goal invokes the Principal Components Analysis (PCA)~\cite{Bhalerao:2014mua}.
We discuss the relation of the partial covariance technique to these methods in App.~\ref{sec:other}.

The outline of our paper is as follows: In Sec.~\ref{sec:partcov} we provide the basic definitions of the partial covariance and partial correlation, and discuss their 
meaning linked to imposing conditions with control variables. Then in Sec.~\ref{sec:fb} we pass to the discussion of the FB multiplicity 
correlations, followed in Sec.~\ref{sec:sup} by the description of the superposition approach to the multi-stage particle production process. Our 
key formulas, allowing to impose centrality fixing conditions at the level of sources rather than hadrons, are derived there. In Sec.~\ref{sec:cf} we 
obtain partial correlations in the initial state for the  BT model, which are confronted with the corresponding results obtained with simulated data in Sec.~\ref{sec:sim}.  
In Sec.~\ref{sec:concl} we recapitulate our results, summarizing in practical terms 
the method based on partial correlations, which can be used in experimental analyses. The appendices contain the discussion
of the relation of the partial covariance analysis to other methods, as well as some technical details.

\section{Partial correlation \label{sec:partcov}}

In this section we establish the notation and provide the standard definitions, with more details 
presented in Appendix~\ref{sec:pcd}.

The simplest case of partial correlations~\cite{Cramer:1946,krzanowski:2000} involves two physical random variables, $X$ and $Y$, and a single control random variable $Z$.   
One defines the element of the covariance matrix in the standard way as
\begin{eqnarray}
\cov{A}{B}&=&\br{AB}-\br{A}\br{B}, \;\;\;\; A,B=X,Y,Z, \nonumber
\end{eqnarray}
where $\br{.}$ denotes the averaging over the sample of $n$ events.
The variance is, of course, the diagonal term, 
\begin{eqnarray}
\var{A}=\br{A^2}-\br{A}^2=\cov{A}{A}.
\end{eqnarray}
The partial covariance between $X$ and $Y$ with the control variable $Z$ 
is defined by the formula (see Appendix~\ref{sec:pcd} for interpretation)
\begin{eqnarray}
\cov{X}{Y\bullet Z} = \cov{X}{Y}-\frac{\cov{X}{Z}\cov{Z}{Y}}{\var{Z}}. \label{eq:covp3}
\end{eqnarray}
The second term in Eq.~(\ref{eq:covp3}) removes the piece of the covariance between $X$ and $Y$ which is 
due to their correlation to $Z$, by means of projecting out the components of $X$ and $Y$ which are parallel to $Z$ (in the $n$-dimensional space, 
with $n$ denoting the number of events). 
The partial variance is, correspondingly,
\begin{eqnarray}
\var{A \bullet Z} = \var{A}-\frac{\cov{A}{Z}^2}{\var{Z}} = \cov{A}{A\bullet Z}, \;\;\; A=X,Y. \nonumber \\  \label{eq:varp3}
\end{eqnarray}

In experimental studies one often uses the correlation function defined as the covariance scaled with the multiplicities, i.e., 
\begin{eqnarray}
C(X,Y)=\frac{\cov{X}{Y}}{\br{X} \br{Y}}, \label{eq:Csc}
\end{eqnarray}
and, correspondingly, $V(A)=\var{A}/\br{A}^2$. Then the partial $C$-correlation following from Eq.~(\ref{eq:covp3}) is 
\begin{eqnarray}
C(X,Y\bullet Z) = C(X,Y)-\frac{C(X,Z)C(Z,Y)}{V(Z)}. \label{eq:Cp3}
\end{eqnarray}

Finally, one defines the partial analog of Pearson's $\rho$-correlation~\cite{Cramer:1946,krzanowski:2000}, 
\begin{eqnarray}
&&\rho(X,Y\bullet Z)=\frac{ \cov{X}{Y\bullet Z}}{\sqrt{\var{X \bullet Z}\var{Y \bullet Z}}} \label{eq:r} \\
&&=\frac{\rho(X,Y)-\rho(X,Z)\rho(Z,Y)}{\sqrt{1-\rho(X,Z)^2}\sqrt{1-\rho(Z,Y)^2}}, \nonumber
\end{eqnarray}
where 
\begin{eqnarray}
\rho(A,B)= \cov{A}{B}/\sqrt{\var{A}\var{B}}.
\end{eqnarray}

As discussed in a greater detail in Appendix~\ref{sec:cond}, the partial covariance, under quite general assumptions~\cite{Lawrance:1976,Baba:2004} which 
are typically fulfilled in ultra-relativistic heavy-ion collisions, is related to the conditional covariance, namely
\begin{eqnarray}
\cov{X}{Y\bullet Z} \simeq \cov{X}{Y\vert Z}, \label{eq:equiv}
\end{eqnarray}
where the meaning of the condition in $\cov{X}{Y\vert Z}$ is that first $Z$ is fixed to a very narrow class 
(for instance, if it describes the discrete multiplicity of hadrons in the reference bin, it can be fixed to a natural number 
equal to this multiplicity), then the covariance 
between $X$ and $Y$ is evaluated within this subsample, and finally averaging of thus obtained covariances over various values of $Z$ within the sample is performed.  

As a matter of fact, 
such a conditional procedure was used in the STAR experiment~\cite{Abelev:2009ag} to analyze the 
FB multiplicity correlations in Au+Au and $p$+$p$  collisions at 
$\sqrt{s_{NN}}=200$~GeV. In this context,  relation~(\ref{eq:equiv}) was derived by Lappi-McLerran~\cite{Lappi:2009vb} under the assumption of 
normal distributions, and by Bzdak~\cite{Bzdak:2011nb} with the condition (\ref{eq:prop}), similarly as in~\cite{Lawrance:1976,Baba:2004}.

The practical significance of Eq.~(\ref{eq:covp3}) or (\ref{eq:Cp3}) is that the imposition of external constraints (such as fixing centrality) may be, under 
general assumptions, accomplished via the partial covariance technique. 

\section{Forward-backward multiplicity correlations \label{sec:fb}}

In our study, $X$, $Y$, and $Z$ random variables are the multiplicities of produced hadrons in, correspondingly, a forward pseudorapidity bin $F$,   
a backward pseudorapidity bin $B$, and a reference bin. We will explore two cases: reference bin $C$ located at a central pseudorapidity bin, and the sum of 
bins $L$ and $R$, located symmetrically at peripheral rapidities $\eta_L$ and $\eta_R=-\eta_L$. Multiplicity in the reference bin determines the centrality of the event. The location of the reference bin
is fixed, whereas $F$, centered around $\eta_1$, and $B$, centered around $\eta_2$, vary in the covered pseudorapidity range (in this more general arrangement we do not 
request that $F$ is forward and $B$ backward, but they assume any location in pseudorapidity within the acceptance range). 
We will also consider the case where the multiplicities in $L$ and $R$ bins 
are fixed independently, according to the case with two constraints described in App.~\ref{sec:pcd}.

Experimental studies of the longitudinal multiplicity correlations have a long history. 
Early investigations of $pp$ and $p\bar{p}$ collisions~\cite{Uhlig:1977dc,Alpgard:1983xp,Alner:1987wb,Aivazian:1988ui,Ansorge:1988fg,Derado:1988ba,Alexopoulos:1995ft} 
and nuclear collisions~\cite{Bachler:1992psr,Akiba:1997yg} were
followed by the studies of ultra-relativistic heavy-ion and $pp$ reactions at RHIC~\cite{Back:2006id,Abelev:2009ag} and at the 
LHC~\cite{ATLAS:2012as,Jia:2015jga,Adam:2015mya,Jia:2016jlg,Aaboud:2016jnr,Sputowska:2016xxx,Adam:2016iwf}. 
Numerous physical models and theoretical methods have been invented in attempts 
to understand the mechanisms behind the generation of long-range correlations~\cite{Capella:1978rg,Kaidalov:1982xe,Chou:1984wp,Capella:1992yb,%
Amelin:1994mf,Braun:2000cc,Giovannini:2002za,Braun:2003fn,Haussler:2006rg,Brogueira:2006yk,%
Armesto:2006bv,Vechernin:2007zza,Braun:2007rf,%
Konchakovski:2008cf,Bzdak:2009dr,Lappi:2009vb,Bzdak:2009xq,Bozek:2010vz,Yan:2010et,deDeus:2010id,Bialas:2011vj,%
Bzdak:2011nb,Bzdak:2012tp,Vechernin:2012bz, De:2013bta,Ma:2014pva,Bzdak:2015dja,Vechernin:2015upa,%
Monnai:2015sca,Broniowski:2015oif,He:2016qwg,He:2016qjs,Olszewski:2016eti}. 
The importance of these investigations lies in the well-known fact that the correlations over a large rapidity separation 
can originate only from the earliest stages of the collision, thus they may reveal fingerprints of the early dynamics of the system.

In our study we demonstrate the methodology based on partial correlations with simulated events for Pb+Pb collisions at $\sqrt{s_{NN}}=2.76$~TeV, 
described in detail in Sec.~\ref{sec:sim}.

\section{Superposition approach \label{sec:sup}}

One should bear in mind that the particle production in ultra-relativistic heavy-ion collisions is an effect of a  
multi-stage process (see, e.g.,~\cite{Olszewski:2013qwa,Olszewski:2015xba}). 
First, we have an early production of entropy, resulting from partonic physics. It is modeled in various approaches, such as string 
formation~\cite{Amelin:1994mf,Brogueira:2006yk}, the Color Glass 
Condensate theory~\cite{Kovner:1995ja,Iancu:2000hn}, or the Glauber 
approach~\cite{glauber1959high,Czyz:1969jg,Bialas:1976ed,Kharzeev:2000ph,Back:2001xy,Broniowski:2007ft}, used in this work.  
The initial partonic entropy is distributed in space in a correlated way. 

This distribution, treated event-by-event, 
serves as an initial condition for the intermediate hydrodynamic evolution (for
reviews see, e.g.,~\cite{Heinz:2013th,Gale:2013da} and references therein) or for 
transport modeling~\cite{Lin:2004en}.
In our work the applied hydrodynamics is deterministic, hence it does not introduce extra 
fluctuations, which should arise in a viscous system~\cite{Kapusta:2011gt}, 
but which according to estimates are not very significant~\cite{Yan:2015lfa,Gavin:2016nir}.

The intermediate evolution continues until freeze-out, where the Cooper-Frye formalism~\cite{Cooper:1974mv} 
is applied at the freeze-out hyper-surface defined with a constant temperature (we take $T_f=150$~MeV).  
It generates {\em primordial} hadrons 
(stable and resonances) according to a thermal distribution. Subsequently, resonances 
undergo decays, possibly in cascades, into stable particles. Due to a statistical nature of the 
production process, the distribution of a hadron of a given species is Poissonian, hence (trivial) 
fluctuations are generated due to the sampling with a finite number of particles.  

The goal of the data analyses is to unfold of the trivial 
fluctuations~\cite{Gazdzicki:1992ri,Fu:2003yf,Bialas:2003gq,Gorenstein:2011vq,Ling:2015yau,Nonaka:2017kko}, 
such as those from statistical hadronization, and acquire
information on correlations generated in the earlier evolution phases. Below we describe how this is accomplished in the 
superposition approach. 

Let us first bring up an important approximation underlying this approach, which may be termed as {\em no bin mixing}. 
The initial distribution of entropy may be divided into cells labeled with their space-time rapidity
\begin{eqnarray}
\eta_\parallel=\frac{1}{2} \log \left ( \frac{t+z}{t-z} \right )
\end{eqnarray}
(and also by the transverse 
coordinates $x$ and $y$). 
The bins are ``carried over'' with hydrodynamics or transport to the freeze-out hyper-surface, with the assumption of no 
mixing between the bins, such that the final pseudorapidity of the fluid element, $\eta$, is a function of $\eta_\parallel$. 
The hydrodynamic push in the longitudinal direction is not very strong, and we estimate 
\begin{eqnarray}
\eta \simeq k \eta_\parallel, \label{eq:etaeta}
\end{eqnarray}
with $k=1.2$
for the model we apply. 

At freeze-out, there is some thermal dispersion of the momenta of hadrons, as particles originating 
from the same hydrodynamic fluid cell acquire rapidities spread with $\Delta \eta \sim 1$. This 
causes some bin mixing of the two-particle correlation function in pseudorapidity. 
In addition, resonance decays generate extra short-range correlations with the width of $\Delta \eta \sim 1$ (we will return to the issue of separating 
the short-range components when describing our results in Sec.~\ref{sec:sim}).

We remark that inclusion of the detector acceptance into the framework does not lead to a new element. 
If the acceptance is expressed with a Bernoulli trial of success rate $p$, then its folding with a Poisson distribution from the thermal motion 
leads to a Poisson distribution with a mean enhanced by the factor $p$. Sums of various particle species with Poisson distributions also lead to 
a Poisson distribution with the mean expressed as a sum means of the added distributions. 

Our basic methodology is as follows: we will use simulated data (with removed short-range component coming from resonance decays) to obtain
the two-particle partial correlation function with the trivial fluctuations unfolded with the help our formalism. The result will be directly compared to the 
partial correlation function of the initial state used in the simulations (cf.~Sec.~\ref{sec:cf}), 
with arguments shifted according to Eq.~\ref{eq:etaeta}.

We now come to the derivation of relevant formulas.
One may consider the fluid cells at the freeze-out hyper-surface as {\em sources}, which emit 
hadrons. Each source, by definition, emits particles independently of other sources, but with the same (for 
the sake of simplicity) distribution. The number of fluid cells (sources) at the forward $F$ and backward $B$
pseudorapidity is denoted as $S_F$ and $S_B$, respectively (recall that the fluid cells at a given rapidity are located at various transverse positions and their number fluctuates).  
Then, the number of produced hadrons in each bin is
\begin{eqnarray}
N_A=\sum^{S_A}_{i=1} m_i,\quad A=F,B
\end{eqnarray}
where $m_i$ denotes the number of particles produced from source $i$. 
We assume that each sources produces particles with the same distribution, hence $\langle m_i \rangle =\langle m \rangle$ and
$\var{m_i}=\var{m}$.
The assumption of the independent production 
leads to simple formulas which connect
moments of the number of sources with the moments of the number of the produced particles~\cite{Olszewski:2013qwa,Olszewski:2015xba}:
\begin{eqnarray}
\br{S_A}\br{m} &=& \br{N_A}, \label{eq:sf}\\
\cov{S_A}{S_{A'}}\br{m}^2 &=&\cov{N_A}{N_{A'}}-\delta^{AA'}\omega(m)\br{N_A}  \nonumber \\
&\equiv&\covov{N_A}{N_{A'}}, \nonumber
\end{eqnarray}
where $\omega(X) =\var{X}/\br{X}^2$ is the scaled variance.
We notice a subtraction of $\omega(m)\br{N_A}$  appearing for the diagonal term $A=A'$, i.e., for the variance:
\begin{eqnarray}
\var{S_A}\br{m}^2 &\equiv&  \cov{S_A}{S_{A}}\br{m}^2 = \var{N_A}-\omega(m)\br{N_A} \nonumber \\ 
&\equiv& \varov{N_A}. \label{eq:supfor}
\end{eqnarray}
The origin of this term is the presence of variance (or autocorrelation) of the particles produced from the same sources, $\var{m_i}=\cov{m_i}{m_i} >0$, 
whereas the covariance of the overlaid distributions from different sources vanishes by the assumption of independent production, $\cov{m_i}{m_j} = 0$ for $i \neq j$.

For the special case of the Poisson distribution of the overlaid variable
$m$ (which is the case of our numerical study presented in Sec.~\ref{sec:sim})), we have $\omega(m)=1$ and $\varov{N_A}=\br{N_A(N_A-1)}-\br{N_A}^2$,
which corresponds to the subtraction of autocorrelations; the average number of pairs appears in the formula.
However, Eq.~(\ref{eq:supfor}) is more general, holding, under the assumptions of the superposition model, for any distribution of the overlaid variable $m$. 
In this paper, we refer to the subtraction of $\omega(m)\br{N_A}$ from the variance as to ``the removal of autocorrelations'',  and indicate it with a bar.

In an experimental study, where in principle one does not know the distribution of $m$, one may find $\omega(m)$ in a numerical way with the following procedure:
One examines the covariance matrix $\cov{N_A}{N_{A'}}$ as a two-dimensional matrix in $A$ and $A'$ indices. The diagonal term at $A=A'$ forms a sharp 
discontinuous ridge, sticking up from a smooth ''background'' function. One then adjusts the value of $\omega(m)$ to remove the sharp ridge from the function 
$\cov{N_A}{N_{A'}}-\delta^{AA'}\omega(m)\br{N_A}$. This prescription conforms to Eq.~(\ref{eq:sf}), as 
the covariance of the sources $\cov{S_A}{S_{A'}}$ is a smooth function of $A$ and $A'$.

The meaning of Eq.~(\ref{eq:supfor}) is that the covariance of the number sources is proportional to the covariance of the observed hadron multiplicities, but with the
autocorrelations removed.
Passing to the scaled covariance~(\ref{eq:Csc}), we have
\begin{eqnarray}
C(S_A,S_{A'})=C(N_A,N_{A'})-\delta^{AA'}\frac{\omega(m)}{\br{N_A}} \equiv \overline{C}(N_A,N_{A'}). \nonumber \\
\end{eqnarray}

We are now ready to build the partial covariance for the superposition approach. Some introductory 
discussion is in place. As stated in Sec.~\ref{sec:partcov}  and further discussed in Appendix~\ref{sec:cond}, the meaning of the partial 
covariance is, essentially, an imposition of a condition on the value of the control variables. In the case of multiplicity correlations in 
ultra-relativistic heavy-ion collisions a first instinct is to constrain the centrality fluctuations, i.e., the fluctuations of number of hadrons 
in a reference bin. However, a more desired constraint concerns the number of sources corresponding to the reference bin. Such a constraint 
is more directly related to a physical situation in the initial state. Suppose a reference bin, defined as $C$, has in a given event $S_C$ sources which ``determine 
the physics''. On the other hand, the number of detected hadrons $N_C$ is sensitive to the fluctuation in the production from sources.
Therefore a fixed 
value of $N_C$ corresponds to event-by event fluctuating values of $S_C$; constraining $N_C$ does not completely constrain $S_C$ and the 
physics of the initial condition remains washed out. 

According to our formalism, the constraint imposed at the level of initial sources is realized with the equation
\begin{eqnarray}
C(S_F,S_B\bullet S_C) &=&  \overline{C}(N_F,N_B)
-\frac{\overline{C}(N_F,N_C)\overline{C}(N_B,N_C)}{\varov{N_C}} \nonumber \\
&\simeq& C(S_F,S_B | S_C).  \label{eq:p1}
\end{eqnarray}
This is the key formula used in our analysis of the simulated data in the following sections.

If one wishes, however, to impose the constraint at the level of the produced hadrons, then using the 
formula 
\begin{eqnarray}
\cov{S_A}{N_{A'}}&=&\cov{S_A}{\sum_i^{S_{A'}}m_i}=\br{m}\cov{S_A}{S_{A'}}\nonumber
\end{eqnarray}
one arrives at the expression
\begin{eqnarray}
C(S_F,S_B\bullet N_C) &=& \overline{C}(N_F,N_B)
-\frac{\overline{C}(N_F,N_C)\overline{C}(N_B,N_C)}{\var{N_C}} \nonumber \\
&\simeq& C(S_F,S_B | N_C).  \label{eq:p2}
\end{eqnarray}
Note that the subtle but important difference between Eq.~(\ref{eq:p1}) and (\ref{eq:p2}) is in the denominator of the subtracted term, where we find the variance 
of the multiplicity in the reference bin with autocorrelations subtracted, or present. In our sample, the autocorrelations increase the variance by $\sim 100$\%, hence the 
effect is very important.

We stress that despite its simplicity, the meaning of Eq.~(\ref{eq:p1}) is non-trivial, as it allows to impose a strict centrality constraint at the level of {\em sources} 
and infer partial correlation of sources, whereas the evaluation is based solely on 
measured multiplicities of the {\em produced hadrons}. 

\section{Modeling initial correlations \label{sec:cf}}

For our illustrative purposes we use the wounded quark model~\cite{Bialas:1977en,Anisovich:1977av} for the initial state.
In this model, the initial sources are the wounded quarks, moving forward or backward, according to the motion of their parent nucleons
from nuclei $A$ and $B$.
An advantage of using the wounded quarks compared to wounded nucleons~\cite{Bialas:1976ed} is that one obtains proper 
scaling~\cite{Eremin:2003qn,KumarNetrakanti:2004ym,Adler:2013aqf,Adare:2015bua,Lacey:2016hqy,Bozek:2016kpf} 
of the multiplicities on the number of participants with no need for the binary-collision component~\cite{Kharzeev:2000ph}.
The event-by-event distribution of wounded quarks  
in the transverse plane is obtained from the Glauber simulations with {\tt GLISSANDO}~\cite{Broniowski:2007nz,Rybczynski:2013yba}, 
corresponding the transverse location of the wounded quarks. The 
longitudinal profile in spatial rapidity $\eta_\parallel$ is taken according to the 
model of ``triangles''~\cite{Bialas:2004su,Adil:2005bb,Bozek:2010bi}, where each source has the entropy distributed  
preferentially in the direction of its motion, according to a simple formula  
\begin{equation}
f_{A,B}(\eta_\parallel)= \frac{y_b \pm \eta_\parallel}{y_{\rm b}} h(\eta_\parallel), \;\;\;\; {\rm for } \ |\eta_\parallel|<y_{\rm b}, \label{eq:lprof}
\end{equation}
where $A$ and $B$ denote the sources belonging to, respectively, the left- and right- moving nuclei,  and 
$y_{\rm b}$ is the rapidity of the beam ($y_{\rm b}\simeq 8$ for Pb+Pb collisions at $\sqrt{s_{NN}}=2.76$~TeV), and 
$ h(\eta_\parallel)$ is an additional profile, typically of a flatten Gaussian form~\cite{Bozek:2010bi}. As $ h(\eta_\parallel)$ cancels 
from the formulas for symmetric $A$+$B$ collisions, we do not need to specify it explicitly.

Let us introduce the notation $Q_A$ and $Q_B$ for the number of wounded quarks belonging to the $A$ and $B$ nuclei, and 
$Q_\pm=Q_A\pm Q_B$. Then, according to Eq.~(\ref{eq:lprof}), the number of sources (combined from nucleus $A$ and $B$) 
at location $\eta_\parallel$ is 
\begin{eqnarray}
S(\eta_\parallel)=\left( Q_+ + Q_- \frac{\eta_\parallel}{y_b}  \right ) h(\eta_\parallel), \label{eq:mBT}
\end{eqnarray}
For symmetric collisions the average over events yields $\br{Q_-}=0$, hence $\br{S(\eta_\parallel)}=\br{Q_+} h(\eta_\parallel)$.

Bzdak and Teaney (BT)~\cite{Bzdak:2012tp} computed the correlation function in the model of triangles (in the variant with wounded nucleons).
It yields a very simple result 
\begin{eqnarray}
C(S_F,S_B)=\frac{\var{Q_+}}{\br{Q_+}^2}+\frac{\var{Q_-}}{\br{Q_+}^2}u_1 u_2, \label{eq:bzdak}
\end{eqnarray}
where 
\begin{eqnarray}
u_{1,2}=\frac{\eta_{\parallel 1,2}}{y_b} = \frac{\eta}{k y_b} \label{eq:u12}
\end{eqnarray}
(cf. Eq.~(\ref{eq:etaeta})), and indices $1$ and $2$ correspond to labels $F$ and $B$, respectively. 
Note that, as announced, the overall rapidity profile $ h(\eta_\parallel)$ cancels out. The moments of $Q_\pm$ are read out from 
{\tt GLISSANDO} simulations via averaging over some chosen class of events.

Next, we use Eq.~(\ref{eq:Cp3}) to derive the partial covariance function
for the BT model with the control bin $C$ placed at $\eta_\parallel=0$. A short calculation yields
\begin{eqnarray}
C(S_F,S_B\bullet S_C)= \frac{\var{Q_-}}{\br{Q_+}^2}u_1 u_2. \label{eq:covpbzd}
\end{eqnarray}
We note that Eq.~(\ref{eq:covpbzd}) differs from Eq.~\ref{eq:bzdak} by not carrying the term with $\var{Q_+}$.
This is clear from the point of view of the conditional correlation, 
as in the present case $C(S_F,S_B| S_C)$ corresponds to fixing the multiplicity at $\eta_\parallel=0$. From 
Eq.~(\ref{eq:mBT}) we get $S(0)=Q_+ h(0)$, hence this is equivalent to fixing $Q_+$, thus  $\var{Q_+}=0$ in the 
calculation of the conditional covariance. The obtained consistency verifies
in an obvious way the relation $C(S_F,S_B\bullet S_C)=C(S_F,S_B| S_C)$.

For the $\rho$-correlation of Eq.~(\ref{eq:r}) we find the very simple formula 
\begin{eqnarray}
\rho(S_F,S_B\bullet S_C)= {\rm sgn}(u_1 u_2), \label{eq:sign}
\end{eqnarray}
where ${\rm sgn}$ denotes the sign function. This means, that the partial $\rho$-correlation of multiplicities in bins located at rapidities of the 
same (opposite) sign is $+1$ ($-1$), indicating maximum correlation (anti-correlation). 

In experiments, it frequently happens that multiplicities in peripheral forward ($R$) and distant backward ($L$) bins are available and can be 
used for centrality determination. Below we consider two cases: 1)~where the sum of the multiplicities in $L$ and $R$ is taken 
and Eq.~(\ref{eq:Cp3}) is used, and 2)~the multiplicities in $L$ and $R$ are taken as separate constraints according to Eq.~(\ref{eq:C4}).
We consider the case where the peripheral bins are symmetrically arranged, with $R$ and $L$ around pseudorapidities $\eta_R$ and $\eta_L=-\eta_R$. 

In case~1) the BT model yields exactly the same result as Eq.~(\ref{eq:covpbzd}), 
\begin{eqnarray}
C(S_F,S_B\bullet S_L+S_R)= \frac{\var{Q_-}}{\br{Q_+}^2}u_1 u_2, \label{eq:LpR}
\end{eqnarray}
which follows from the fact that according to Eq.~(\ref{eq:mBT}) $S(\eta_L)+ S(\eta_R) \sim{Q_+}$, and 
the condition with symmetrically arranged peripheral bins fixes $Q_+$, as in the case of the central bin. 

In case~2) Eq.~(\ref{eq:C4}) gives for the BT model a vanishing result, 
\begin{eqnarray}
C(S_F,S_B\bullet S_L,S_R)=0, \label{eq:LR}
\end{eqnarray}
which is compatible with the simultaneous constraints $S(\eta_L)=0$ and $S(\eta_R)=0$, which fixes both $Q_+$ and $Q_-$, hence 
$\var{Q+}=\var{Q_-}=0$, and $C(S_F,S_B | S_L,S_R)=0$.
We note that this is a specific feature of the BT model, whereas models which include extra fluctuations in the initial state, e.g., 
the fluctuating-string model of~\cite{Broniowski:2015oif}, would yield a non-zero result.

\begin{figure*}[tb]
\begin{center}
\includegraphics[width=0.497\textwidth]{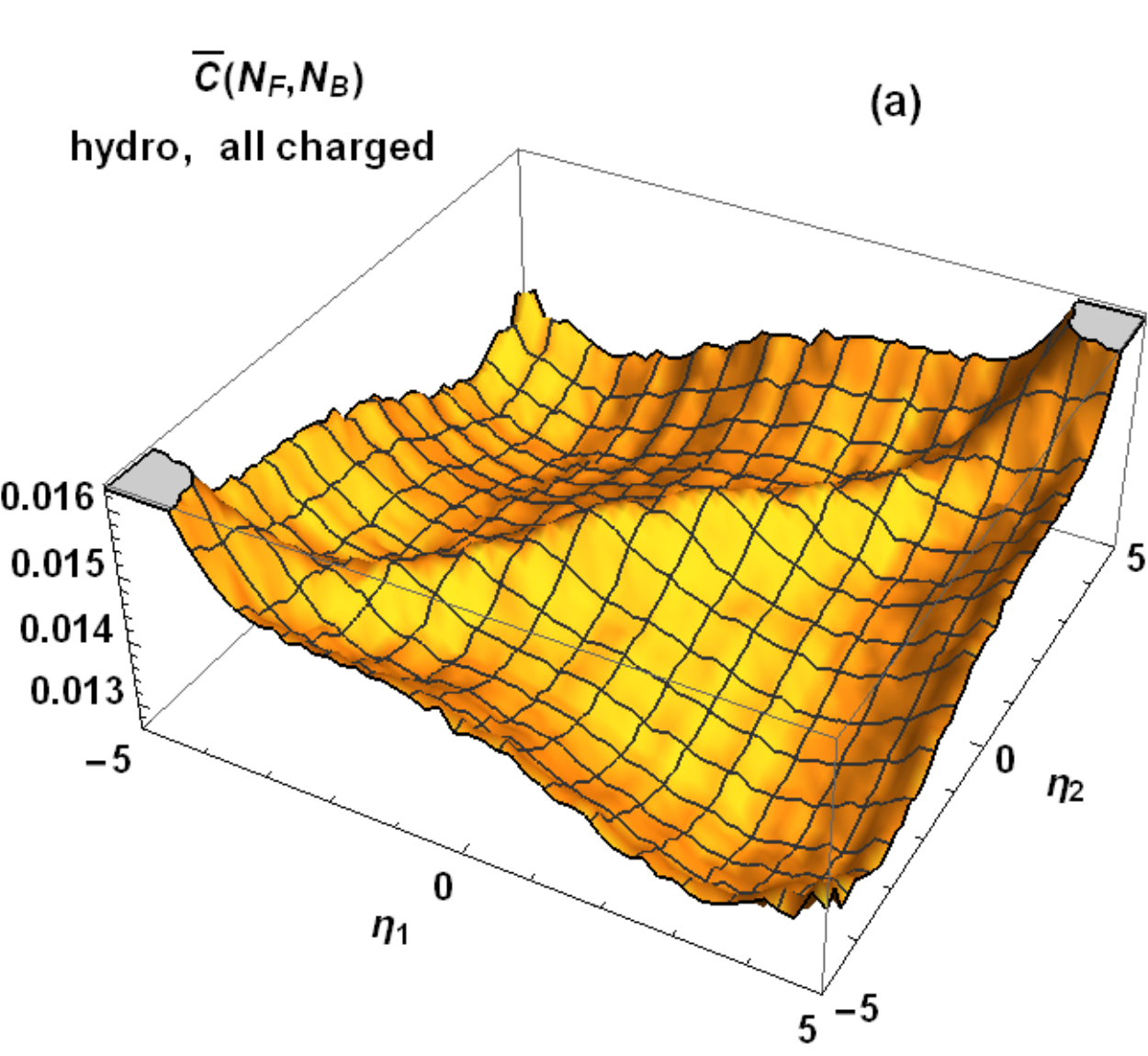}\hfill \includegraphics[width=0.497\textwidth]{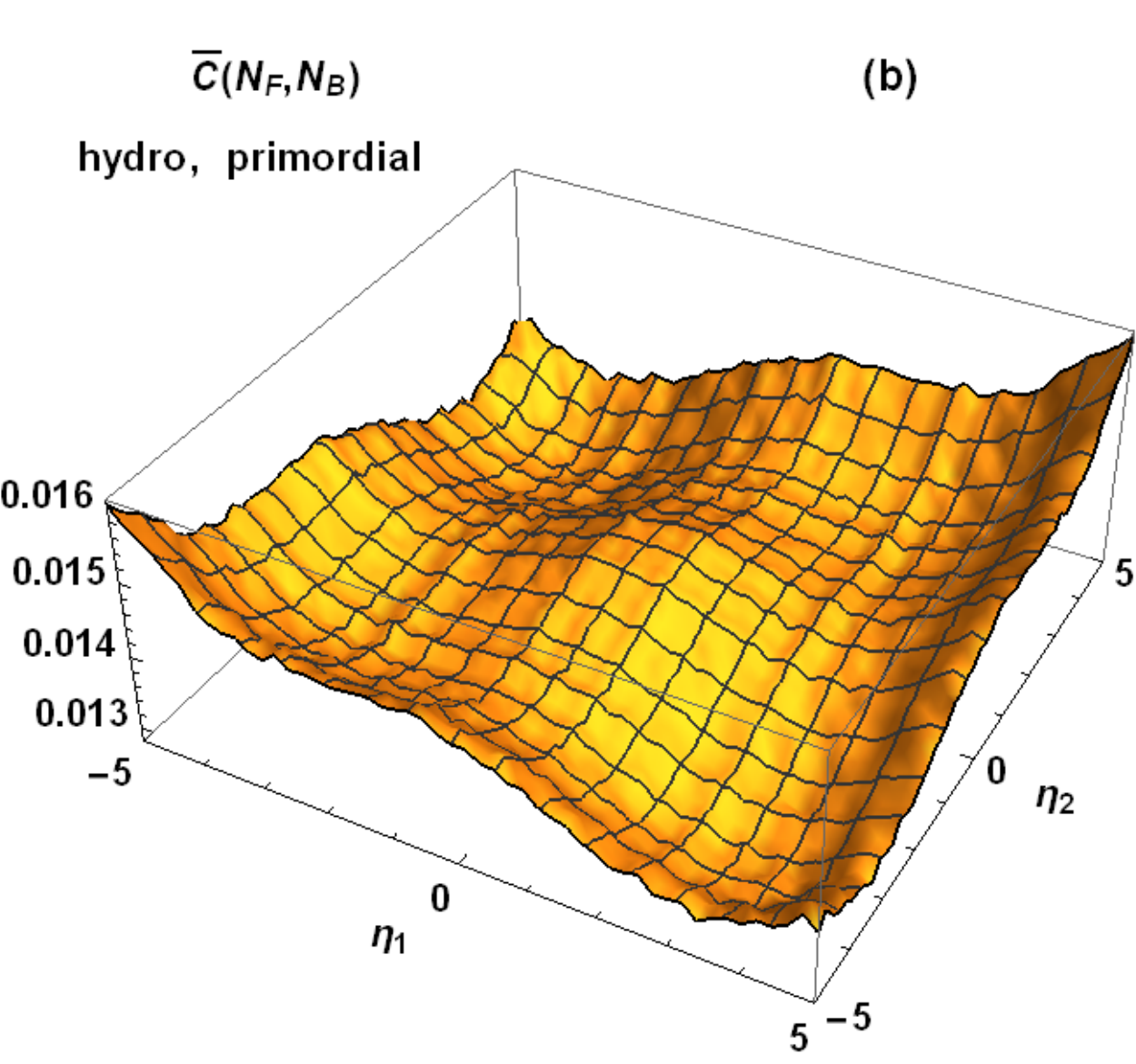}
\caption{Two-particle $\overline{C}$-correlation (with removed autocorrelations) in pseudorapidity of hadron multiplicities 
for the simulated data (a) for all charged particles after resonance decays, and  
(b)~for the primordial hadrons.  Wounded quark model, 3+1D viscous event-by-event hydrodynamics, statistical hadronization. 
Pb+Pb collisions at  $\sqrt{s_{NN}}=2.76\textrm{ TeV}$ in centrality class 30-40\%. \label{fig:cov}}
\end{center}
\end{figure*}

\section{Results for the simulated events \label{sec:sim}}

Our final goal is to test to what extent the formulas obtained in the previous section are
reproduced with simulated data and the master formula~(\ref{eq:p1}) and its equivalent for the case of 
left and right peripheral bins.  Because of departures from assumptions of the superposition approach (bin mixing, resonance decays)  
this is not an academic exercise, but a verification if the method may be practical in actual data analyses.

\begin{figure*}[tb]
\begin{center}
\includegraphics[width=0.497\textwidth]{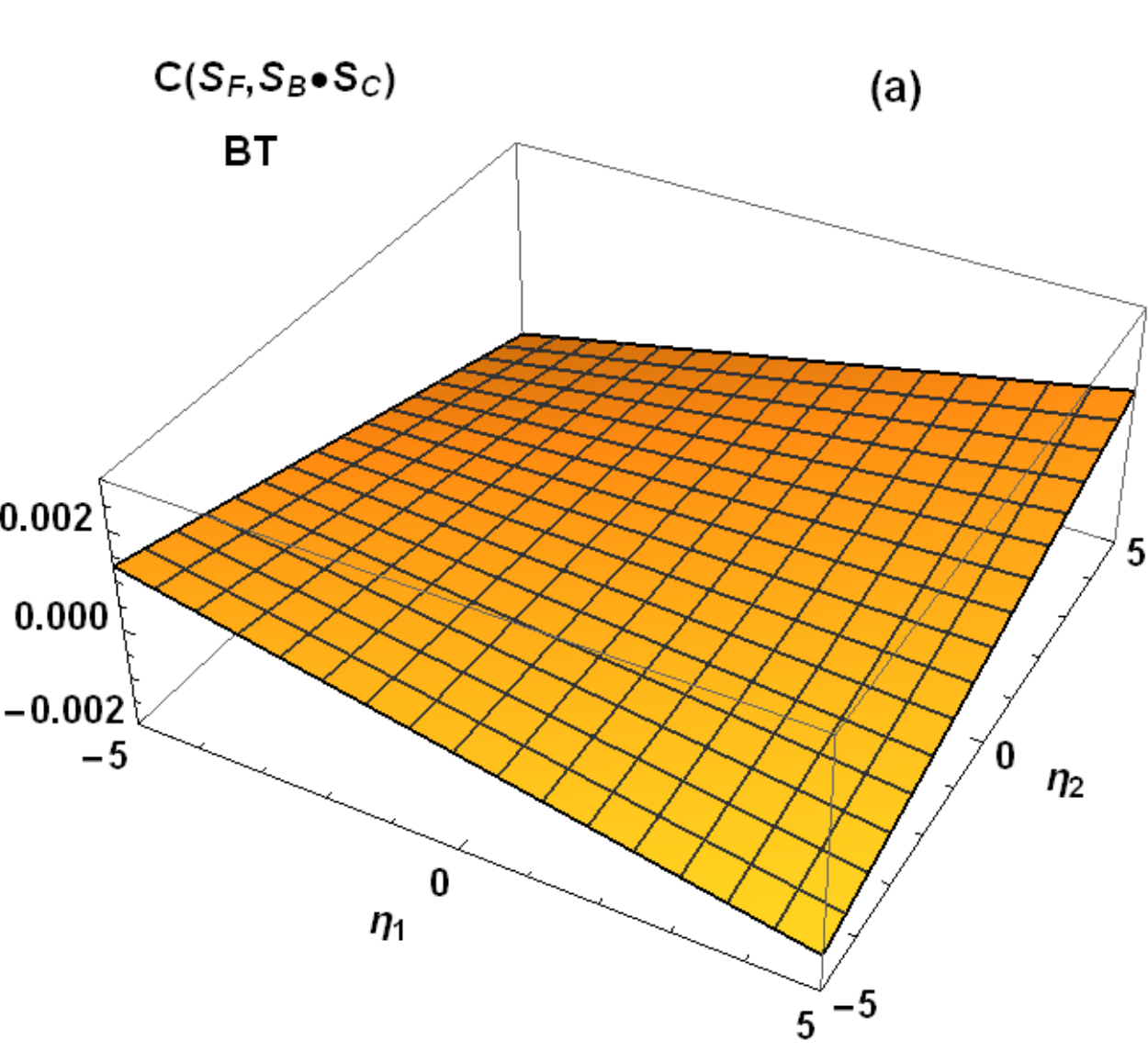} \hfill \includegraphics[width=0.497\textwidth]{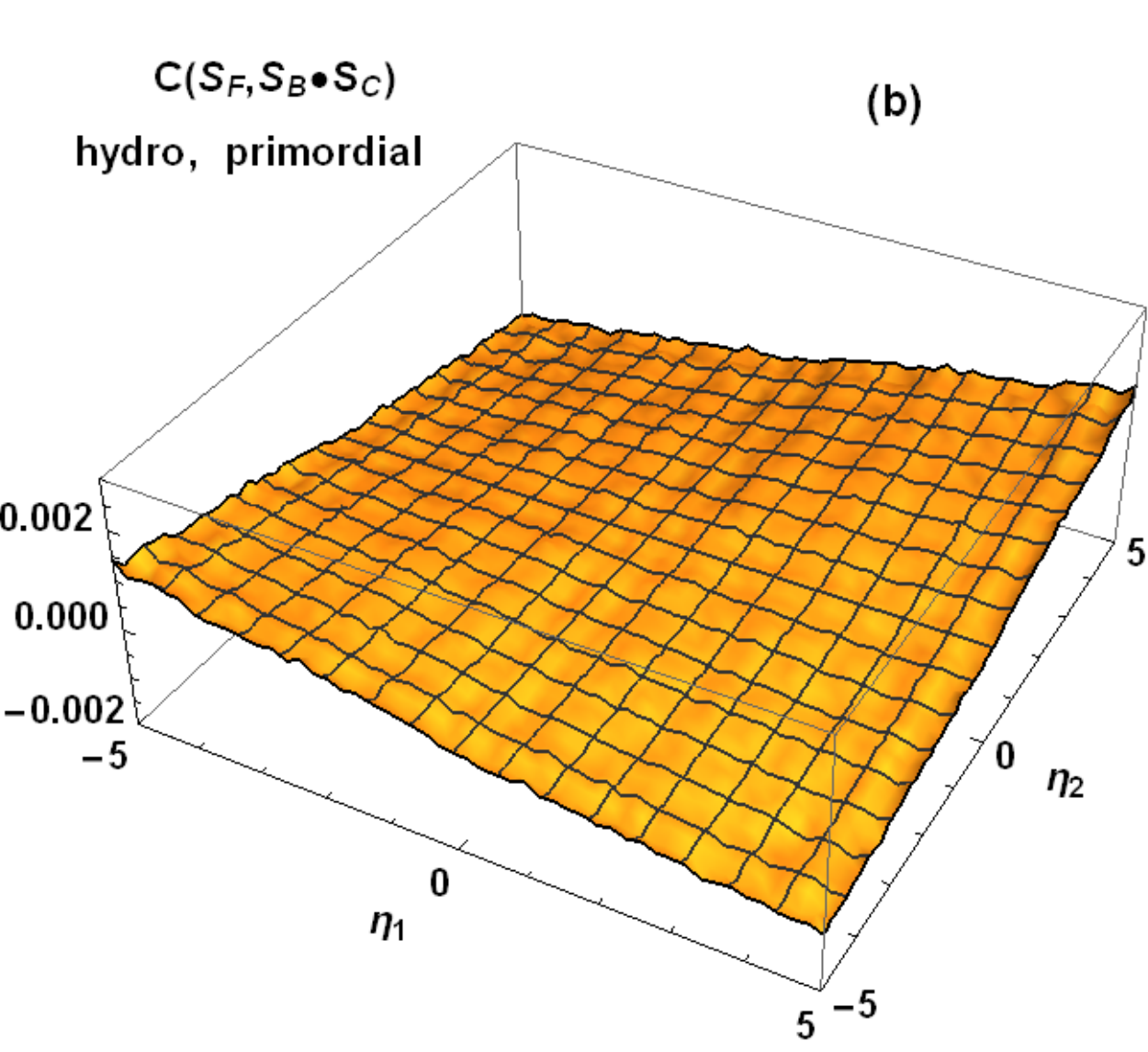}
\caption{Partial $C$-correlation function for the source multiplicities with the central reference bin (a) for the BT model, and  
(b)~for the simulated data with primordial hadrons.
\label{fig:pcov}}
\end{center}
\end{figure*}

The simulated data are obtained as follows: for the initial condition, we apply the wounded quark model 
for Pb+Pb collisions at $\sqrt{s_{NN}}=2.76\textrm{ TeV}$, as described in 
Sec.~\ref{sec:cf}. 
We take a rather broad sample corresponding to centrality (as determined with $Q_+$) of 30-40\%. Then 
we use the results of event-by-event 3+1D viscous hydrodynamics~\cite{Bozek:2009dw} obtained with the wounded-quark initial conditions. 
The statistical hadronization at freeze-out is carried out with {\tt THERMINATOR}~\cite{Kisiel:2005hn,Chojnacki:2011hb}, which incorporates all 
hadrons from the Particle Data Tables and implements resonance decays.
We label the results obtained with the products of resonance decays as ``all charged'', which consists of $\pi^\pm$, $K^\pm$, $p$, 
and $\bar{p}$.  To reduce the correlations induced by the resonance decays, we also present the results obtained 
with the ``primordial'' particles generated at freeze-out, i.e., before the resonance decays.
 
As we wish to have a view as broad  as possible, we take a large continuous 
acceptance window in our numerical study, with $|\eta_{1,2}|\le 5.1$ divided into 51 {\em physical} bins of width of 0.2. Our coverage is 
thus (on purpose) much larger than accessible in the LHC experiments, to better illustrate the approach. The central {\em reference} bin, where the number of 
hadrons in $N_C$, is taken as $\eta_C \le 0.5$, thus it has a width of five physical bins (other values of width of the reference bin could be taken, with similar 
results up to statistical uncertainties).

\begin{figure*}[tb]
\begin{center}
\includegraphics[width=0.497\textwidth]{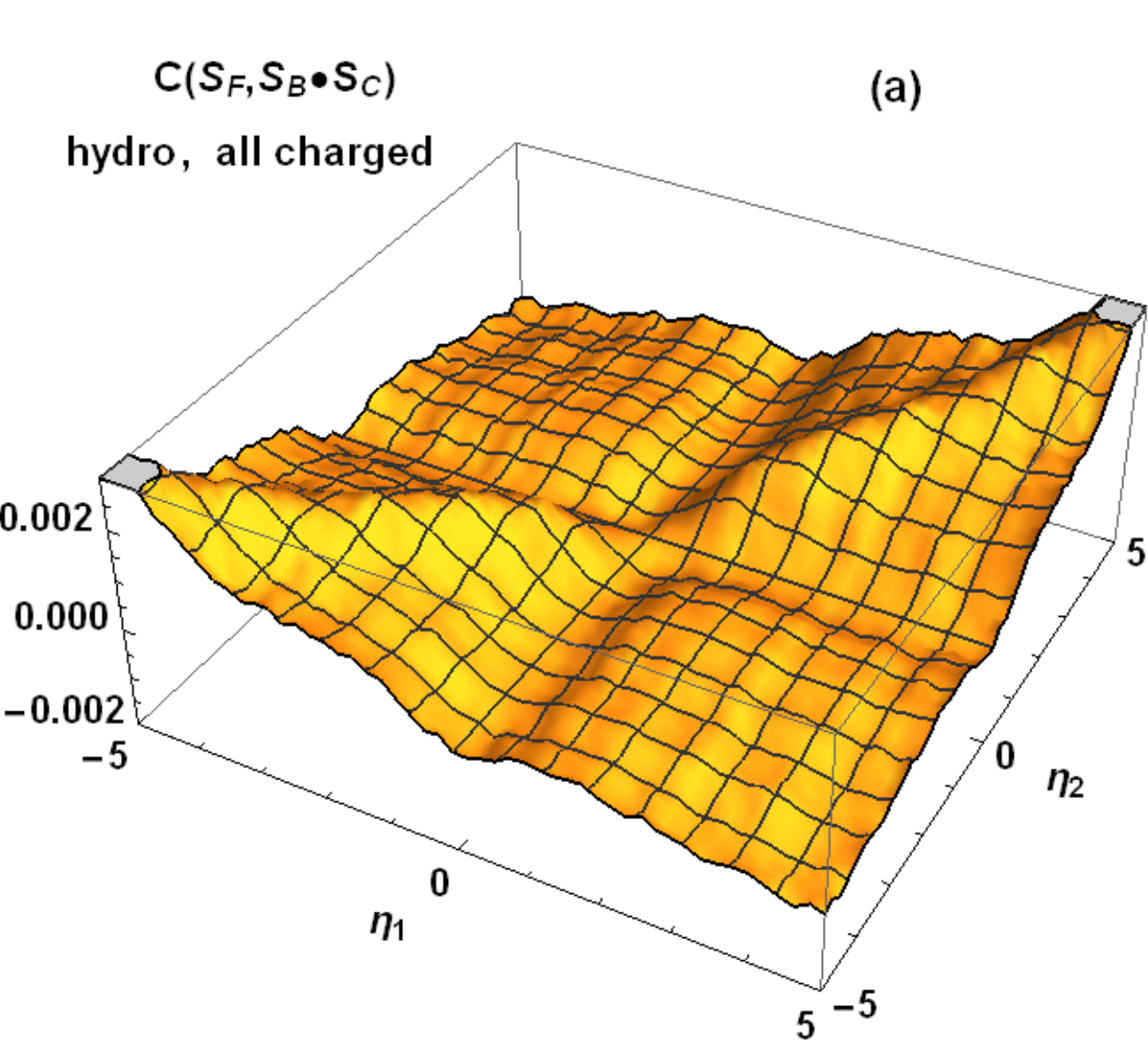} \hfill \includegraphics[width=0.497\textwidth]{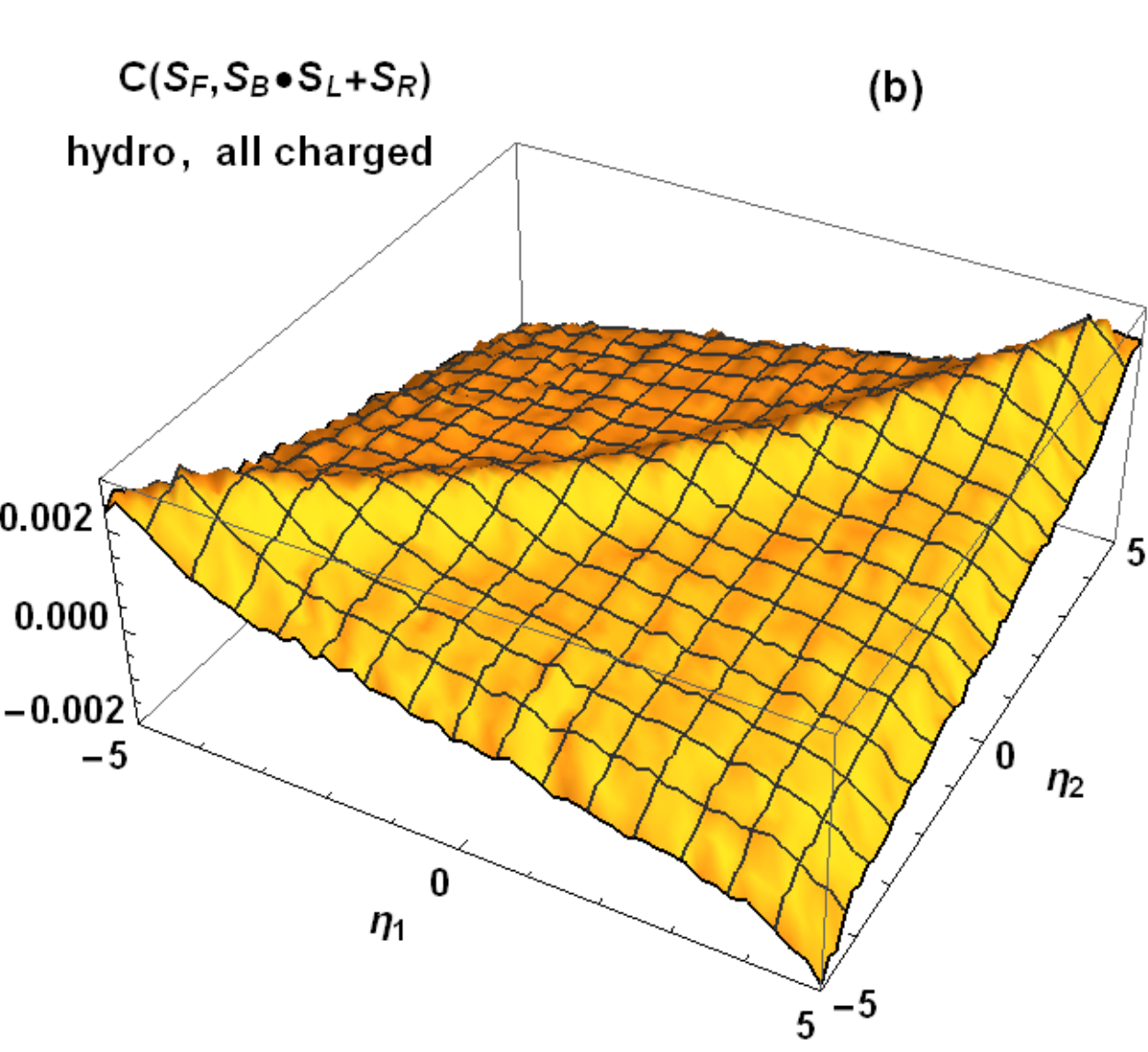}
\caption{Partial $C$-correlation function for the source multiplicities for the simulated data for all charged hadrons, obtained
with (a)~the central reference bin, and (b)~with the sum of left and right peripheral bins. 
\label{fig:CLR}}
\end{center}
\end{figure*}

Figure~\ref{fig:cov} presents the basic output from the simulated data, namely the $\overline{C}$-correlation 
(with autocorrelations removed), obtained with (a)~all charged particles (i.e., after 
resonance decays), and (b)~for the primordial particles. In case (a) we note a hallmark ridge, wide by about one unit of pseudorapidity, extending along the diagonal. 
It is due to the resonance decays, which is the basic 
difference between cases (a) and (b).  We also 
note other non-trivial structures in the $\overline{C}$-correlation arising from the applied hydrodynamic model, such as its rise at the boundaries, however 
understanding these intricate details is not the goal of this work. Rather, we take the simulated data (which, as we see, are not trivial) as they are, 
and then carry out the partial correlation analysis outlined in the previous sections.

In Fig.~\ref{fig:pcov} we compare the partial $C$-correlation function
for the source multiplicities with the central reference bin, obtained for the BT model with Eq.~(\ref{eq:covpbzd}), shown in panel~(a),
to the same quantity obtained from the simulated data with {\em primordial} hadrons with Eq.~(\ref{eq:p1}).
We note that the overall agreement of~(a) and~(b) is quite remarkable.  Hence, if we had the possibility of separating the resonance decays
(and in real data, also other sources of short-range correlations from later stages of the evolution) we could infer the initial state correlations with 
the presented methodology. Note that the application of Eq.~(\ref{eq:p1}) leads to large cancellations when passing from Fig.~\ref{fig:cov}(b) to 
Fig.~\ref{fig:pcov}(b). Also, the non-uniformities are ``miraculously'' smoothed out, hence they had to originate from centrality fluctuations.

Separating other sources of correlations is of course far from  simple. One method (as done by the ATLAS collaboration in~\cite{ATLAS:anm}), is to do 
a numerical fit to a function describing the short-range corrections, and then simply subtract it. In our illustration, it would correspond to ``skimming''
the ridge in Fig.~\ref{fig:cov}(a), and then carrying out the calculation, which would lead to a result similar to Fig.~\ref{fig:pcov}(b). 

We note that our model does not account for the correlations reflecting the conservation laws, which is not essential for testing the partial covariance approach.

\begin{figure*}[tb]
\begin{center}
\includegraphics[width=0.497\textwidth]{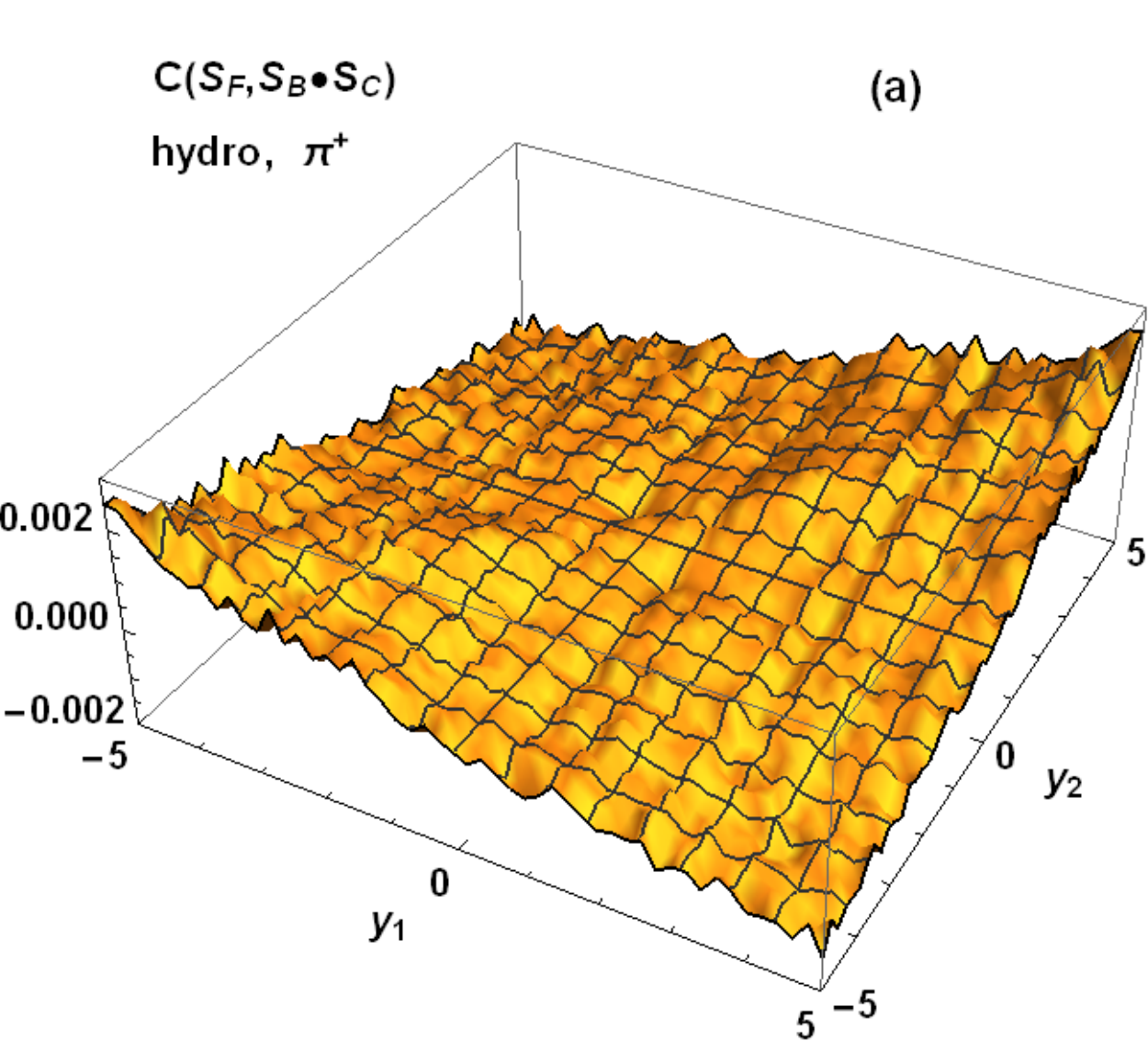} \hfill \includegraphics[width=0.497\textwidth]{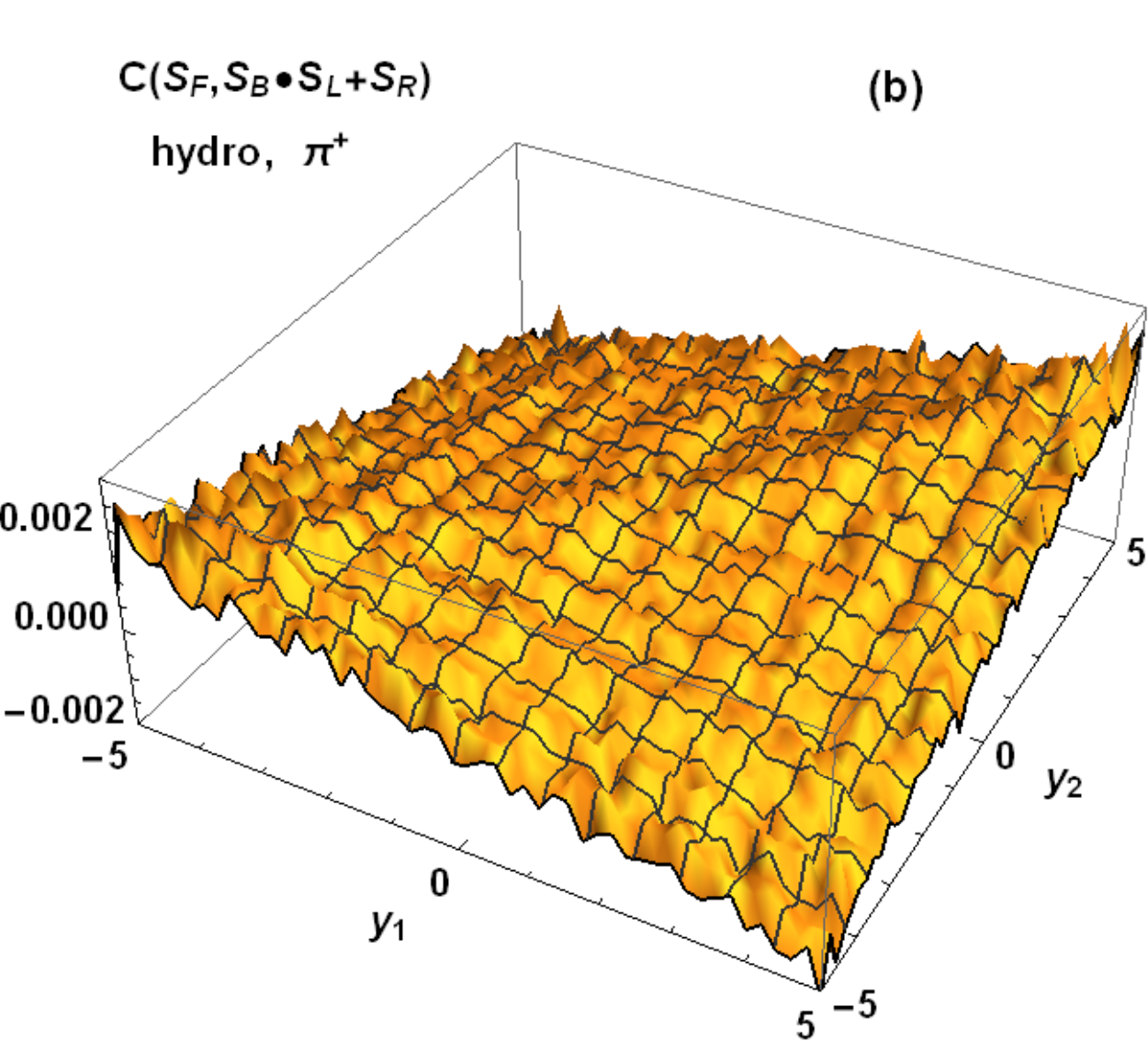}
\caption{Partial $C$-correlation function for the source multiplicities for the simulated data for $\pi^+$,
with (a)~the central reference bin, and (b)~with the sum of left and right peripheral bins.
\label{fig:CLRpi}}
\end{center}
\end{figure*}

\begin{figure*}[tb]
\begin{center}
\includegraphics[width=0.497\textwidth]{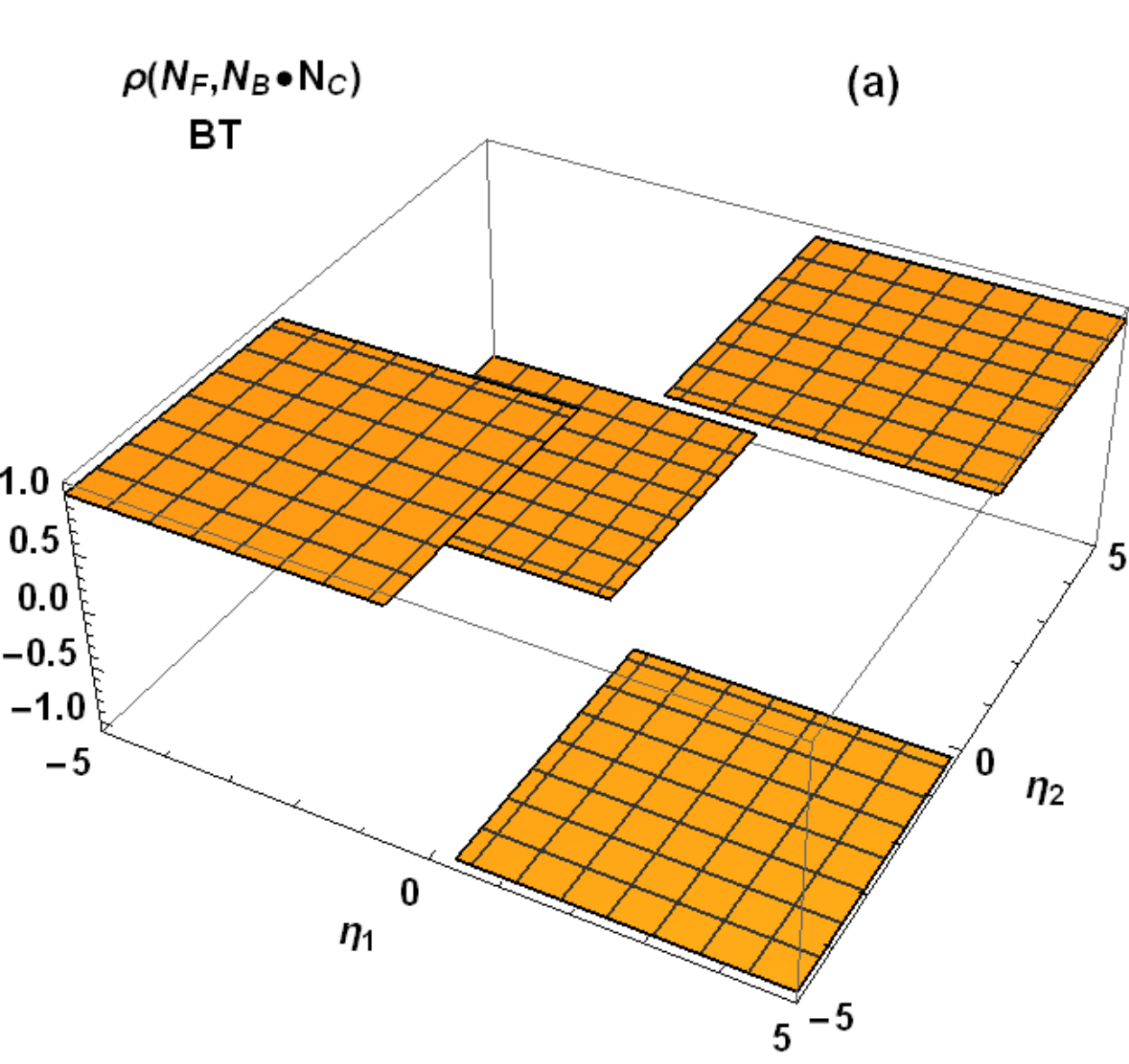} \hfill \includegraphics[width=0.497\textwidth]{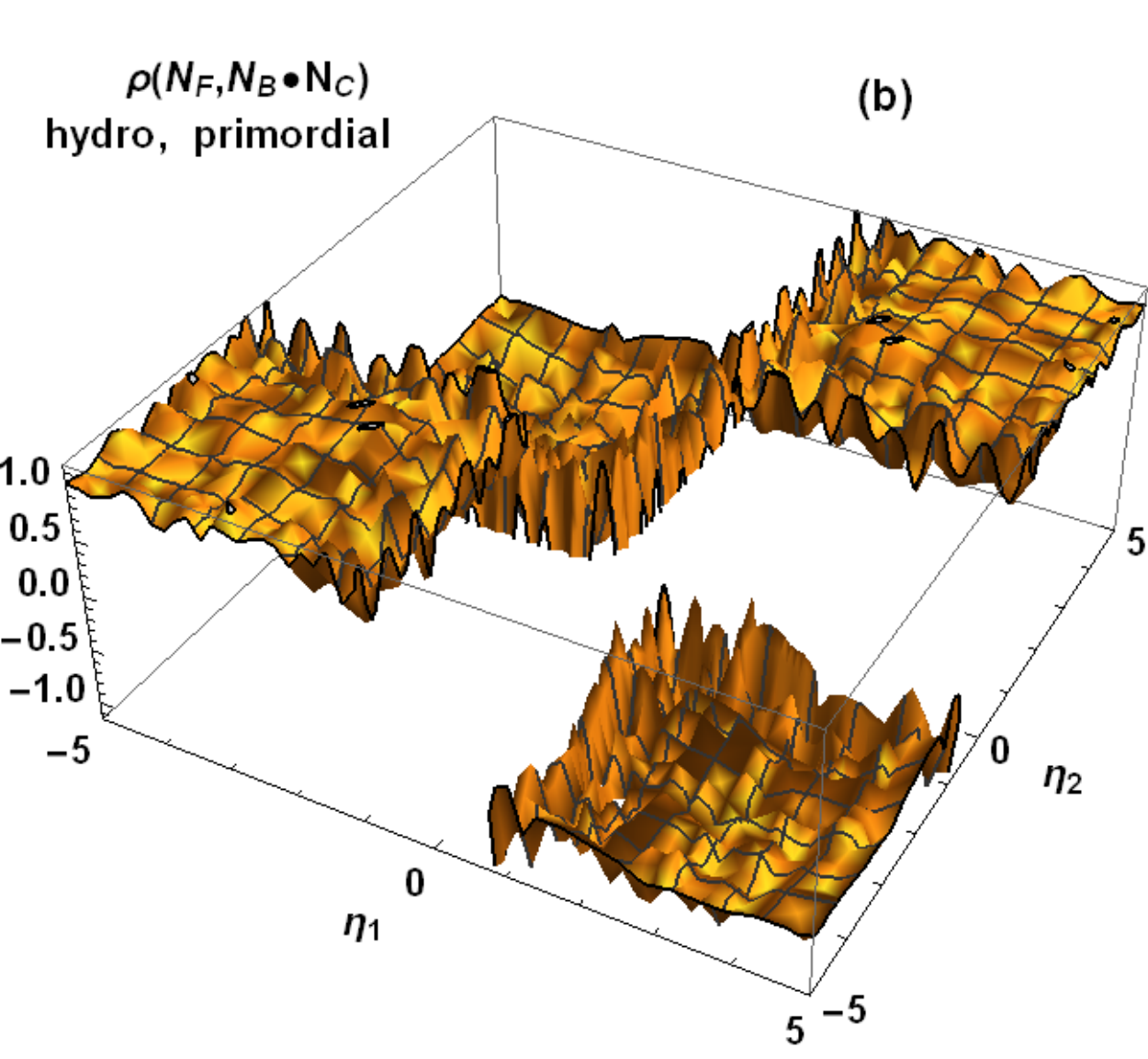}
\caption{Partial $\rho$-correlation for the source multiplicities (a) for the BT model, and  
(b) for the simulated data with primordial hadrons. 
\label{fig:prho}}
\end{center}
\end{figure*}

When the short-range effects from the resonance decays are kept, i.e., we are using the data from Fig.~\ref{fig:cov}(a), then the resulting partial $C$-correlation with the central 
reference bin has the shape shown in Fig.~\ref{fig:CLR}(a). We note the ridge along the $\eta_1=\eta_2$ diagonal, but also several other features. First, we can see a depletion 
along the lines $\eta_1=0$ and $\eta_2=0$. This is a simple artifact of the central reference bin placed at $\eta=0$, since the definition~(\ref{eq:Cp3}) has the feature, that 
when one of the measurement bins is equal to the reference bin, e.g., $Y=Z$, then $C(X,Z\bullet Z)=0$ identically. Suppose we decompose the measured correlation 
function into the short- ($s$) and long-range ($l$) components, $\overline{C}=\overline{C}_{s}+ \overline{C}_{l}$. Then Eq.~(\ref{eq:p1}) becomes
\begin{eqnarray}
&& C(S_F,S_B\bullet S_C) =  \overline{C}_{s}(N_F,N_B)+ \overline{C}_{l}(N_F,N_B) - \label{eq:srlr} \\
&& \frac{[\overline{C}_{s}(N_F,N_C)\!+\!\overline{C}_{l}(N_F,N_C)] [\overline{C}_{s}(N_B,N_C)\!+\!\overline{C}_{l}(N_B,N_C)]}
{\overline{\rm v}_{s}(N_C)+\overline{\rm v}_{l}(N_C)}. \nonumber
\end{eqnarray}
We see from here all the features coming out in Fig.~\ref{fig:CLR}(a). When $F\simeq B$, we get the ridge from $\overline{C}_{s}(N_F,N_B)$, and when 
$F\simeq C$ or $B\simeq C$, the function drops to zero for the reason discussed above Eq.~(\ref{eq:srlr}). Note, however, that even when $F$, $B$, and $C$ are
all sufficiently separated, we get still an artifact from the presence of ${\rm v}_{s}$ in the denominator of the second term in Eq.~(\ref{eq:srlr}). It leads to a reduction 
of the subtraction, and, consequently, larger values of $C(S_F,S_B\bullet S_C)$ in Fig.~\ref{fig:CLR}(a) (in the region where $F$, $B$, and $C$ are separated) than in the 
BT model shown in Fig.~\ref{fig:pcov}(a).

Similar distortions are seen in the case of the $L+R$ reference bin, displayed in Fig.~\ref{fig:CLR}(b). We note that $C(S_F,S_B\bullet S_L+S_R)$ is 
``pulled down'' at the peripheries, and too large in the region where $F$, $B$, $L$ and $R$ are well separated.

The conclusion of the above discussion is that the short-range components must be separated at the level of $\overline{C}(N_F,N_B)$ for the 
presented analysis to make practical sense.

\begin{figure*}[tb]
\begin{center}
\includegraphics[width=0.497\textwidth]{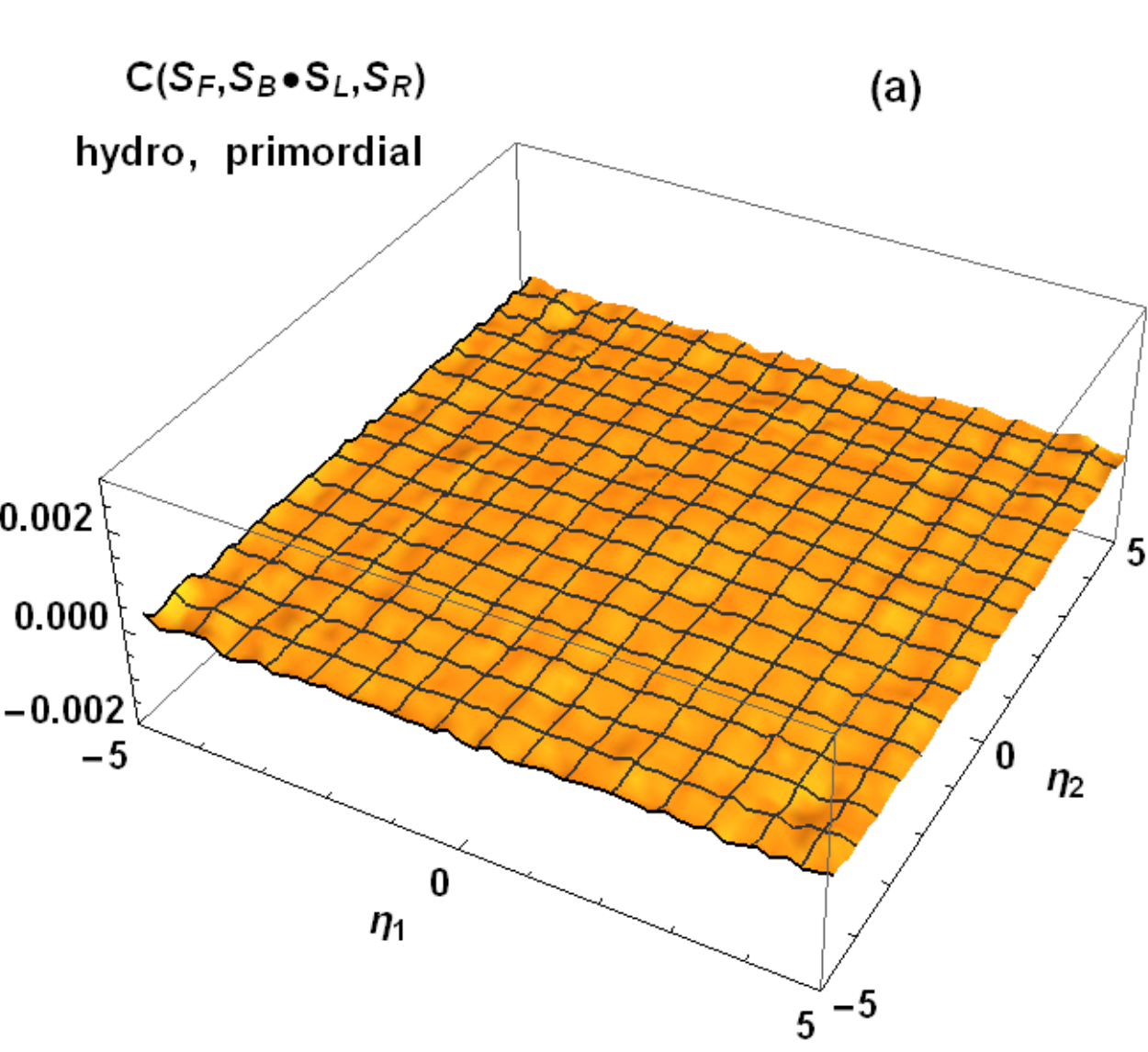} \hfill \includegraphics[width=0.497\textwidth]{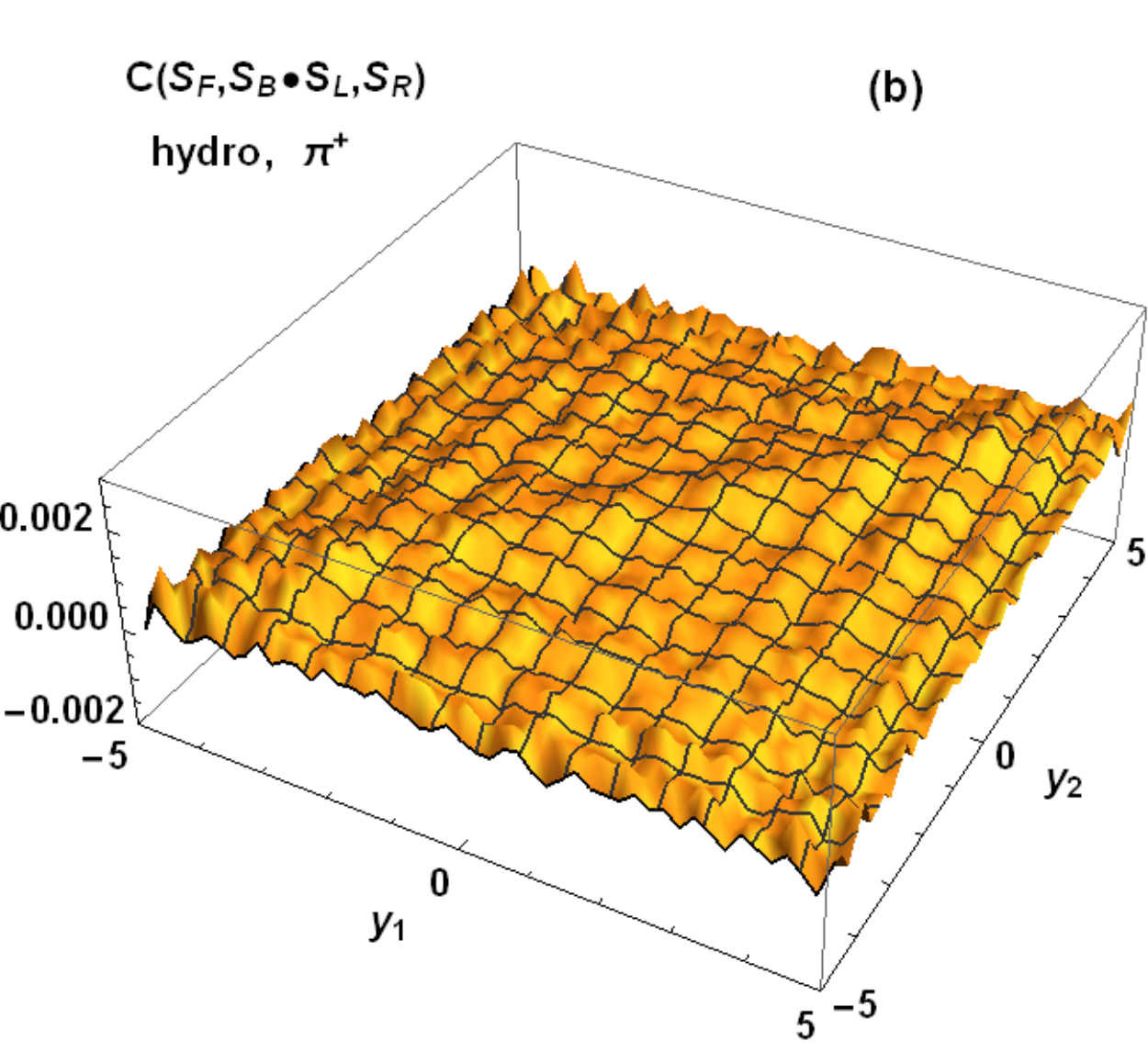}
\caption{Partial $C$-correlation function for the source multiplicities with two peripheral reference bins, obtained for the simulated data 
with (a)~primordial hadrons, and (b)~all positively charged pions, $\pi^+$. 
\label{fig:LR}}
\end{center}
\end{figure*}

A well-known method of reducing correlations from the resonance decays is to use particles of the same charge. We apply our procedures for 
$\pi^+$, as there are no resonances that decay into $\pi^+ \pi^+$ pairs. Only some remnant correlation is expected from resonance decays proceeding
in cascades. The result is shown in Fig.~\ref{fig:CLRpi}, with the central reference bin in panel (a), and sum of the peripheral bins $L+R$ in panel (b). 
We note a very weak correlation from cascade decays, visible along the $y_1=y_2$ diagonal. In the region away from the diagonal we see a 
very close agreement with Fig.~\ref{fig:pcov}. Therefore the use of hadrons of the same sign is an efficient way of getting rid of the short-range correlations 
from the resonance decays. 

In Fig.~\ref{fig:prho} we show the partial $\rho$-correlation of Eq.~\ref{eq:r}, with the central reference bin. We recall that in the BT model 
it is given by Eq.~(\ref{eq:sign}), visualized in panel (a). As expected, this shape is reproduced within statistical noise by the primordial particles  
of the simulated data. We remark that the $\rho$-correlation, being the ratio of the covariance and variance, carries less information than the 
$C$-correlation. From the $C$-correlation we can read off both the scaled covariance and variance (along the diagonal).

Finally, we test the partial correlation approach for the case of the two peripheral reference bins discussed in App.~\ref{sec:pcd}, with  $-6.1<\eta_L<-5.1$ and 
$5.1<\eta_R<6.1$. We recall that in the BT model the corresponding partial correlation vanishes, cf.~Eq.~(\ref{eq:LR}). This is also the case, 
within numerical uncertainties,
for the simulated data, as visualized in Fig.~\ref{fig:LR}, for the case of primordial particles (a) and all positively charged pions (b).

\section{Conclusions \label{sec:concl}}

We have presented a simple method capable of providing information on the initial two-particle multiplicity correlations, which is insensitive to 
centrality fluctuations.  The basic formalism 
relies on the concept of partial covariance, where a reference bin (or a few reference bins) are used to impose constraints on the sample. Application 
of the method to a superposition model, where particle production occurs in subsequent stages, allows to unfold the trivial statistical  fluctuations from the hadronization 
at freeze-out and gain insight into the correlations in the initial stage of the reaction.

We have demonstrated the feasibility of the method by carrying out an illustrative analysis on simulated data 
obtained with hydrodynamics, run event-by-event on initial conditions provided by the wounded quark model, and followed by 
statistical hadronization. We have shown that performing the calculations for the partial correlations of hadrons which do not carry
correlations from resonance decays (primordial hadrons, or pions of the same charge) reproduces efficiently the initial 
partial correlations. Thus the method can be used as a practical tool in experimental data analysis of two-particle correlations. 

A nontrivial  aspect of our approach is a simple way of unfolding trivial statistical fluctuations generated at statistical hadronization. 
In the superposition model, it amounts to removal of autocorrelations from the building blocks of the partial correlation function. 
Moreover, that way we are able to impose constraints on the number of sources in the reference bin, rather than the number of hadrons, 
which is desirable from the point of view of studying the initial state. 

The method is directly extendable to imposition of more constraints, related to a possible involvement of more detectors. This allows
for getting more information of the correlations generated in the initial state. 

A crucial element of the correlation analyses is the separation of the short-range component, expected to be generated in later stages of the collision, 
and the long-range component, generated in the initial phase. We have demonstrated on the simulated data that the use of same-charge pions 
largely reduces the correlations due to resonances. Other sources of short-range correlations (jets, femtoscopy) should be removed with a suitable 
method.

To summarize, the procedure of obtaining the partial correlations in the initial state is as follows:
\begin{enumerate}
 \item Obtain the two-particle correlation function in (pseudo)rapidity from the data.
 \item Remove autocorrelations and the short-distance component.
 \item Apply Eq.~(\ref{eq:p1}), or its generalizations in the case of more control bins.
 \item Up to corrections from bin mixing and spatial rapidity -- pseudorapidity mapping, the 
 result represents the correlations in the initial state of the collision with centrality fluctuations removed 
 at the level of sources.
\end{enumerate}

Finally, we note that the technique of partial correlations is applicable to other observables which correlate to the ``centrality'' determination, for instance 
various charges or the transverse momentum.

\bigskip
\bigskip

\begin{acknowledgments}
We are grateful to Piotr Bo\.zek for providing us a sample of the hydrodynamic simulation results for the wounded quark model, 
used for illustration of the data analysis methods discussed in this paper.
This research was supported by the Polish National Science Centre grant 2015/19/B/ST2/00937.
\end{acknowledgments}

\begin{appendix}

\section{Partial covariance with more control variables \label{sec:pcd}} 

In this appendix we list some basic definitions and properties referring to the partial correlations.
In a general case we have $n$ physical random variables
\mbox{${X}=\bn{X_1,\ldots,X_n}$} and  $m$ control random variables \mbox{${Z}=\bn{Z_1,\ldots,Z_m}$}. The 
quantities $X_i$ and $Z_j$ are vectors in the $N_{\rm ev}$-dimensional space, where $N_{\rm ev}$ (the number of events) is the 
number of the data points. Averaging over events for a quantity $A$ is defined as $\langle A \rangle = 1/N_{\rm ev} \sum_{k=1}^{N_{\rm ev}} A_i$.
One defines the partial covariance matrix as
\begin{eqnarray}
\Sigma_{{XX}\bullet {Z}}=\Sigma_{{XX}}-\Sigma_{{XZ}}\Sigma^{-1}_{{ZZ}}\Sigma_{{ZX}}, \label{eq:covnb}
\end{eqnarray}
where $\Sigma_{{AB}}$ is the standard covariance matrix defined as
\begin{eqnarray}
\left( \Sigma_{{AB}} \right )_{ij} &=&  \langle (A_i - \langle A_i \rangle ) ( B_j -  \langle B_j \rangle) \rangle  \nonumber \\ &=& \langle A_i B_j \rangle - \langle A_i \rangle \langle B_j \rangle, \; A,B=X,Z, 
\end{eqnarray}
where $i$ and $j$ label the variable types.
Mathematically, Eq.~(\ref{eq:covnb}) corresponds to projecting out from the vectors ${X}$ the components belonging to the space spanned by  the vectors ${Z}$ 
(shifted to their central values). 
Indeed, introducing the projected physical vectors 
\begin{eqnarray}
\overline{X}_i=X_i - \left ( \Sigma_{{XZ}} \right )_{ij} \left ( \Sigma^{-1}_{{ZZ}} \right )_{jj'} \left( Z_{j'} - \langle Z_{jj'} \rangle \right),
\end{eqnarray}
which by construction gives the orthogonality condition 
\begin{eqnarray}
\langle Z_m  \overline{X}_i \rangle - \langle Z_m \rangle \langle  \overline{X}_i \rangle =0, 
\end{eqnarray}
we straightforwardly verify Eq.~(\ref{eq:covnb}).

The above formulas simplify when the control vectors $Z_j - \langle Z_j \rangle$ are orthonormal (which we can always achieve via the Gramm-Schmidt 
procedure or by diagonalization). 
Denoting these orthonormal vectors as $U_j \equiv Z_j - \langle Z_j \rangle$, $j=1,\dots,m$, we have 
\begin{eqnarray}
\overline{X}_i=X_i - \left ( \Sigma_{{XU}} \right )_{ij} U_j, 
\end{eqnarray}
and
\begin{eqnarray}
\hspace{-1cm} \left ( \Sigma_{{XX}\bullet {Z}} \right )_{ii'} =
\left ( \Sigma_{XX} \right )_{ii'}  - \sum_{j=1}^m \left ( \Sigma_{{XU}} \right )_{ij} \left ( \Sigma_{{UX}} \right )_{ji'}. \label{eq:ortho}
\end{eqnarray}
From construction, the diagonal terms are non-negative, $\left ( \Sigma_{{XX}\bullet {Z}} \right )_{ii} \ge 0$, where the equality occurs when $X_i - \langle X_i \rangle$ 
is contained in the space spanned by the vectors $U_k$.

The interpretation of Eq.~(\ref{eq:ortho}) is straightforward: we subtract from the covariance of the physical variables the covariance proceeding via a correlation, one-by-one, to the 
control variables $U_j$. The meaning of the general case (\ref{eq:covnb}) is the same, with the complication arising from the non-orthonormality of the control variables.   
Some further mathematical facts, in particular the connection to the {\em linear regression} analysis, may be found in~\cite{wiki:partial}. 

Throughout the paper we use the short-hand notation 
\begin{eqnarray}
\hspace{-7mm} c(X_i,X_i  \bullet {Z}) \equiv \left ( \Sigma_{{XX}\bullet {Z}} \right )_{ii'}, \, c(X_i,X_i ) \equiv \left ( \Sigma_{{XX}} \right )_{ii'}.
\end{eqnarray}
The partial covariance matrix scaled with the multiplicities is defined as 
\begin{eqnarray}
C(X_i,X_j  \bullet {Z})=\frac{c(X_i,X_j  \bullet {Z})}{\br{X_i}\br{X_j}}. \nonumber \\
\end{eqnarray}

The Pearson-like partial correlation coefficient between $X_i$ and $X_j$ is defined as 
\begin{eqnarray}
\rho(X_i,X_j  \bullet {Z})=\frac{c(X_i,X_j  \bullet {Z})}{\sqrt{c(X_i,X_i  \bullet {Z})c(X_j,X_j  \bullet {Z})}}, \nonumber \\
\end{eqnarray}
which makes sense as long as $c(X_i,X_j  \bullet {Z})>0$.
From the Schwartz inequality $-1 \le \rho(X_i,X_j  \bullet {Z}) \le 1$, and for the diagonal terms $\rho(X_i,X_i  \bullet {Z})=1$, as in the case of the standard Pearson's 
correlation $\rho(X_i,X_i)$.

For the simplest case of a single control variable, the above formulas reduce to Eqs.~(\ref{eq:covp3}) and (\ref{eq:r}).
For the special case of two physical and two control variables, explored in this paper, we have
\begin{widetext}
\begin{eqnarray}
& &\cov{X}{Y\bullet Z_1,Z_2} = \cov{X}{Y}  \label{eq:cov4b} \\
&&-\frac{\cov{X}{Z_1} \var{Z_2} \cov{Z_1}{Y}+\cov{X}{Z_2}\var{Z_1}\cov{Z_2}{Y}-
\cov{Z_1}{Z_2} [\cov{X}{Z_1}\cov{Z_2}{Y}+\cov{X}{Z_2}\cov{Z_1}{Y}]}{\var{Z_1}\var{Z_2}-\cov{Z_1}{Z_2}^2}.\nonumber
\end{eqnarray}
For the case of the scaled covariance Eq.~(\ref{eq:cov4b}) becomes
\begin{eqnarray}
& &\co{X}{Y\bullet Z_1,Z_2} = \co{X}{Y}-\label{eq:C4} \\
&&\frac{\co{X}{Z_1} V(Z_2) \co{Z_1}{Y}+\co{X}{Z_2}V(Z_1)\co{Z_2}{Y}-
\co{Z_1}{Z_2} [ \co{X}{Z_1}\co{Z_2}{Y}+\co{X}{Z_2}\co{Z_1}{Y}]}{V({Z_1})V({Z_2})+\co{Z_1}{Z_2}^2},\nonumber 
\end{eqnarray}
where $V(Z_j)=\var{Z_j}/\br{Z_j}^2 = C(Z_j,Z_j)/\br{Z_j}^2$. 

For the partial correlation coefficient the explicit formula reads
\begin{eqnarray}
\rho(X, Y\bullet Z_1,Z_2) &=& \frac{A(X,Y)}{\sqrt{A(X,X)A(Y,Y)}}, \\
A(X,Y)&=&\left[1-\rho \left(Z_1,Z_2\right)^2\right] \rho (X,Y)+\rho \left(Z_1,Z_2\right) \left [ \rho \left(Z_2,X\right) \rho \left(Z_1,Y\right)+\rho \left(Z_1,X\right) \rho \left(Z_2,Y\right)\right ] \nonumber \\
 &-&\rho \left(Z_1,X\right) \rho \left(Z_1,Y\right)- \rho \left(Z_2,X\right) \rho \left(Z_2,Y\right). \nonumber
\end{eqnarray}   
\end{widetext}

\section{Relation of the partial covariance to the conditional covariance \label{sec:cond}} 

Lawrance~\cite{Lawrance:1976} has shown that if a sample satisfies the affine condition
\begin{eqnarray}
E\bn{{X}\vert {Y}}=\alpha+{BY}, \label{eq:prop}
\end{eqnarray}
with $\alpha$ a constant and ${B}$ a constant matrix, then the 
equality of the partial covariance and the conditional covariance follows, 
\begin{eqnarray}
\Sigma_{{XX}\bullet {Y}}=\Sigma_{{XX}\vert {Y}}.\label{eq:covcond}
\end{eqnarray}
The converse was shown by Baba, Shibata, and Sibuya~\cite{Baba:2004}, hence conditions (\ref{eq:prop}) and (\ref{eq:covcond}) are equivalent.
The question if  (\ref{eq:prop}) holds, can be tested on the actual data. In the context of ultra-relativistic heavy-ion collisions, it should hold for sufficiently 
narrow centrality classes~\cite{Bzdak:2011nb}. Also, condition (\ref{eq:prop}) holds for normal distributions~\cite{Baba:2004}. 

The important practical implication of Eq.~(\ref{eq:covcond}) is that the partial covariance method is a simple way of imposing constraints in the correlation 
analysis.

\section{Relation to other methods \label{sec:other}}

One may relate the partial covariance method to the Principal Components Analysis (PCA)~\cite{Bhalerao:2014mua}.
Suppose all bins are the {\em measurement} bins $X$, obtained, for instance, by dividing the full pseudorapidity acceptance of the detector info 
narrower bins. Let $U_j$ denote the eigenmodes of the covariance matrix $\Sigma_{XX}$. Then one may project out the eigenmodes (with the highest eigenvalues)
from the covariance matrix, according to Eq.~(\ref{eq:ortho}). Thus, PCA may be viewed as a special case of the Partial Covariance 
method, where the constraints have the form of the eigenmodes of the covariance matrix. 

Whereas the algebra (projection) in the two methods is the same, the accents are different. In PCA one does not separate the measurement and reference variables, treating all the bins 
``democratically" and focuses on a possible hierarchy in the eigenvalues (large gaps in the spectrum are linked to strong correlations, or collectivity, in the fluctuations). 
In the partial covariance method the measured variables are grouped from the outset into 
the {\em physical} bins $X$ and the {\em reference} bins $Y$. This is more natural when the reference data come from different detectors
(e.g.,  in ultra-relativistic heavy-ion collisions, from the peripheral detectors, transverse energy colorimeters, etc.) 
than the detector collecting the physical data (central TPC). In this case the constraints come entirely from the reference variables, and the constraint 
vectors $U_j$ are built from the vectors $Y$ (see App.~\ref{sec:pcd}).

In PCA applied to pseudorapidity fluctuations, the multiplicity eigenmode with a highest eigenvalue, denoted in Ref.~\cite{Bhalerao:2014mua}as $v_0^{(1)}$,
corresponds to an $\eta$-independent fluctuation, i.e., where multiplicities in all the bins vary by an equal amount. This may be interpreted as {\em centrality}
fluctuation, hence the removal of the highest eigenvalue mode in PCA is equivalent to getting rid of the centrality fluctuations, with centrality defined via the total 
multiplicity from all the bins.

The issue of separating centrality fluctuations is also focal the methods aiming at {\em strongly-intensive} fluctuation
measures~\cite{Gazdzicki:1992ri,Gorenstein:2011vq,Sangaline:2015bma,Broniowski:2017tjq}. The framework applied there is also based
on the superposition model, but there is one common source (with its  multiplicity traditionally termed as the ``volume'') and
two types of particles, $A$ and $B$, emitted from the source with multiplicities $m_A$ and $m_B$, respectively. 
One may then use the superposition approach to relate the statistical moments of $m_A$ and $m_B$ to the corresponding moments of 
the observed multiplicities of $A$ and $B$, $N_A$ and $N_B$, in such a way the fluctuation of the sources cancels out. Thus the object of the study is 
the emission process from the source, and not the fluctuation of the number of sources (volume), considered trivial.

With the method described in this paper we study the case where there are multiple {\em types} of sources (each in a given rapidity bin) and (for simplicity) 
just one type of hadrons (an extension to more types of hadrons is possible). 
Our objective is the fluctuation of the number of sources (at various rapidities), whereas the fluctuations of the overlaid variable $m$ are 
considered not interesting. Thus, the partial covariance method used with the superposition approach in this work
is, in a sense, complementary to the studies based on the strongly-intensive fluctuation measures.

\section{The $a_{nm}$ coefficients \label{sec:anm}}

In this appendix we discuss the partial covariance for the case, where the $\overline{C}$-correlation function for hadrons
(with autocorrelations removed) is expressed via the expansion~\cite{Bzdak:2012tp}
\begin{eqnarray}
\overline{C}(\eta_1,\eta_2) = \sum_{m,n=0}^\infty a_{nm} T_n \left(\frac{\eta_1}{Y}\right) T_m\left(\frac{\eta_2}{Y}\right), \label{eq:exp}
\end{eqnarray}
where $T_n(x)$ is a set of orthonormal functions and $[-Y,Y]$ is the pseudorapidity domain.  
The choice of~\cite{ATLAS:2015kla,ATLAS:anm,Jia:2015jga} is 
\begin{eqnarray}
T_n(x)=\sqrt{2+1/2}P_n(x),
\end{eqnarray}
where $P_n(x)$ denote the Legendre polynomials. The orthonormality condition 
$\int_{-1}^1 dx T_n(x) T_m(x)=\delta_{nm}$ is satisfied.
The $a_{nm}$ coefficients are 
\begin{eqnarray}
a_{nm} &=& \int_{-Y}^Y \frac{d \eta_1}{Y} \int_{-Y}^Y \frac{d \eta_2}{Y} \overline{C}(\eta_1,\eta_2)
T_n\left(\frac{\eta_1}{Y}\right) T_m\left(\frac{\eta_2}{Y}\right). \nonumber \\
\end{eqnarray}

The transformation to the partial $C$-correlation function of Eq.~(\ref{eq:p1})
leads, in general, to mixing of the $a_{nm}$ coefficients, i.e., the coefficients $a^C_{nm}$for $C(S_F,S_B\bullet S_C)$ become
complicated functions of the $a_{nm}$ coefficients for $\overline{C}(N_F,N_B)$.

We can, however, derive a simple formula connecting these coefficients,  introducing the expansion (for a fixed $\eta_C$
of the reference bin) 
\begin{eqnarray}
\overline{C}(\eta_1,\eta_C) = \sum_{m}^\infty a_{m}(\eta_C) T_m\left(\frac{\eta_2}{Y}\right), \label{eq:exp2}
\end{eqnarray}
Then 
\begin{eqnarray}
 a^C_{nm}=a_{nm}-\frac{a_n(\eta_C) a_m(\eta_C)}{\overline{\rm v}(\eta_C)} \label{eq:aa}
\end{eqnarray}
(up to the rescaling effects of Eq.~(\ref{eq:etaeta}), not included explicitly in the present discussion).

We remark that for the special case of the BT model with the central reference bin, $a^C_{11}=a_{11}$.

\end{appendix}

\bibliography{hydr}

\end{document}